\documentstyle[11pt]{article}
\newcommand{\be}{\begin{equation}}
\newcommand{\ee}{\end{equation}} %\indent}
\newcommand{\eei}{\end{equation}\indent\indent}
\newcommand{\bc}{\begin{center}}
\newcommand{\ec}{\end{center}}
\newcommand{\ber}{\begin{eqnarray}}
\newcommand{\ear}{\end{eqnarray}}
\newcommand{\ba}{\begin{array}}
\newcommand{\ea}{\end{array}}

\newcommand{\1}{^{(1)}}

\def\case#1/#2{\textstyle\frac{#1}{#2} }
\begin{document}
\begin{titlepage}
\hoffset=-37pt
\title{A Renormalization Group Approach\\
to\\ Relativistic Cosmology}
%\\

\author{{\sc Mauro Carfora}
\\
\normalsize{\it S.I.S.S.A. - I.S.A.S, Via Beirut 2-4, 34014 Trieste,
Italy.}
\\
\normalsize{\it Istituto Nazionale di Fisica Nucleare, Sezione di
Pavia, Via Bassi 6, I-27100 Pavia, Italy.}
\\
{\sc Kamilla Piotrkowska}
\\
\normalsize{\it S.I.S.S.A. - I.S.A.S. via Beirut 2-4, 34014 Trieste,
Italy.}
\\ \\ %its address
                      } %its address
\date{$\mbox{}$ \vspace*{0.3truecm} \\ \normalsize{\today}           }
%%%%%%%%%%%%%%%%%%%%%%%%%%%%%%%%%%%%%%%%%%%%%%%%%%%%%%%%%%%%%%%%%%%%%%
\maketitle
\thispagestyle{empty}
%\vspace*{1truecm}
\vspace*{0.5truecm}
\begin{abstract}
We discuss the averaging hypothesis tacitly assumed in standard
cosmology.  Our approach is implemented in a ``3+1" formalism and
invokes the coarse graining arguments, provided and supported by the
real-space Renormalization Group (RG) methods, in parallel with
lattice models of Statistical Mechanics.  Block variables are
introduced and the recursion relations written down explicitly
enabling us to characterize the corresponding RG flow. To leading
order, the RG flow is provided by the Ricci-Hamilton equations studied
in connection with the geometry of three-manifolds. The possible
relevance of the Ricci-Hamilton flow in implementing the averaging in
cosmology has been previously advocated, but the physical motivations
behind this suggestion were not clear. The RG interpretation provides
us with such physical motivations. The properties of the
Ricci-Hamilton flow make it possible to study a critical behaviour of
cosmological models. This criticality is discussed and it is argued
that it may be related to the formation of sheet-like structures in
the universe.  We provide an explicit expression for the renormalized
Hubble constant and for the scale dependence of the matter
distribution.  It is shown that the Hubble constant is affected by
non-trivial scale dependent shear terms, while the spatial anisotropy
of the metric influences significantly the scale-dependence of the
matter distribution.
\end{abstract}
\centerline{SISSA Ref. 8/95/FM (Jan 95), (Improved Version)}
\end{titlepage}
%%%%%%%%%%%%%%%%%%%%%%%%%%%%%%%%%%%%%%%%%%%%%%%%%%%%%%%%%%%%%%%%%%%%
%
% AFTER THIS, PUT THE VARIOUS SECTIONS
%
%%%%%%%%%%%%%%%%%%%%%%%%%%%%%%%%%%%%%%%%%%%%%%%%%%%%%%%%%%%%%%%%%%%

%%%%%%%%%%%%%%%%%%%%%%%%%%%%%%%%%%A%%%%%%%%%%%%%%%%%%%%%%%%%%%%%%%%%%A
\tableofcontents
\vfill\eject
\section{Introduction}

A successful physical theory usually enables us to isolate some
limited range of length scales, or select a not too big set of
variables, to render the problem tractable and at the same time
preserve its essence.  Fortunately, in many circumstances it is not
necessary to resolve the details associated with each scale, since
generally phenomena of each size can be treated independently. For
example, in hydrodynamics there is no need to specify the motion of
each water molecule and yet waves can be described as a disturbance of
a continuous fluid.

There are also problems that have many various scales of length, that
is to say, phenomena or processes where each length-scale's
contribution is of equal importance. To handle them one has to take
into account the entire spectrum of length scales, dealing with
fluctuations of almost any wavelength and consequently many coupled
degrees of freedom.  For example, critical phenomena, turbulent flow,
the internal structure of elementary particles and confinement in QCD
belong to the above class of problems \cite{Kenneth}.

We argue here that also in gravitational physics we encounter a
problem of a similar nature, namely, the so-called averaging problem
in relativistic cosmology.

\vskip 0.5 cm

Relativistic cosmologies rely on some form of Cosmological Principle.
The latter is usually a smoothing-out hypothesis imposed {\it a
priori} on the distribution of matter in the universe. A well known
example is provided by the Friedmann-Lema\^\i tre-Robertson-Walker
(FLRW) metric which is assumed to describe the real universe.

There are at least two reasons why one should be very careful in doing
so.

Firstly, observations of the distribution of matter have shown that
the scale at which the background homogeneity is reached,\footnote{If
this scale has been reached at all is still a matter of controversy.}
is probably of the order of hundreds of Mpc, meaning that locally,
i.e. at least up to this scale, the universe is quite lumpy.
Consequently, the local geometry is very complex and its nature not
very illuminating from the point of view of cosmology. Seen this way,
the approach usually taken is to average out all the matter, i.e. to
redistribute them in the form of a homogeneous perfect fluid and use
continuous functions (e.g. matter density, pressure) in modeling the
universe, assuming that they represent ``volume averages" of the
corresponding fine scale quantities.  In doing so one tacitly assumes
that such a smoothed-out universe and the real locally inhomogeneous
one, behave identically under their own gravitation. However, it has
been stressed in \cite{bi:ellis} that the above assumption, though
usually taken for granted, is by no means justified. First of all,
Einstein's equations are highly non-linear which is why any averaging
process on them\footnote{Moreover, additional care is required since a
volume average of a tensor is not a covariant quantity, unlike
scalars.} is far from trivial in general.

Secondly, and this is a remark of a philosophical flavour, most of the
observational data are theory dependent, i.e. their meaning can be
interpreted only by assuming a particular theoretical explanation.

Thus it is of great importance that the very foundations of a
cosmological model should be as sound as possible, in particular, they
should be free from assumptions which may not be warranted.
\bigskip

\section{Coarse-graining in Cosmology}

A possible solution to the averaging problem would be to explicitly
construct a procedure for carrying out the smoothing process in the
full theory.  Almost all existing attempts were concerned with the
linearized theory, with a possible exception of
\cite{Mauro} (see also \cite{ICarJ}), for a review see \cite{Kras}.

In \cite{Mauro} a covariant smoothing-out procedure was put forward
for the space-times associated with gravitational configurations which
may be considered near to the standard ones, generating closed FLRW
universes. The procedure makes use of Hamilton's theorem about smooth
deformations of three-metrics and is adapted for smoothing-out an
initial data set for cosmological solutions to the Einstein equations.
While interesting in its own right, this approach seemed rather {\it
ad hoc} and not yet capable of resolving the issues of actual limits
of validity of the FLRW models in cosmology.

\bigskip

Now, our hope is that there is a smart and simpler way to the heart of
the problem, borrowing from the known theories and methods of
statistical mechanics, based on the real-space Renormalization Group
(RG) approach to study critical phenomena in lattice models
\cite{Kadanoff,
Wilson}.  Although the usual renormalization transformations, invoking
averaging over square blocks, are designed mostly having ferromagnetic
systems in mind, there are many more problems suitable for RG methods.
These are difficult problems where the reductionist approach fails and
where the effective degrees of freedom of a physical system are scale
dependent.
%kam
The difficulty of this kind of issues can be traced to a multiplicity
of scales and, moreover, there can presumably be a gross mismatch
between the largest and smallest scales in the problem.  The averaging
problem in cosmology can be looked at and studied as belonging to
precisely this kind of problems. This is our main objective in this
paper.

Some form of RG is active on any system where there are fluctuations
present (they by no means do need to be quantum). This is so, since
one can integrate the fluctuations out of the physical quantities of
interest, e.g. the partition function, and depending on the ``scale"
up to which one is integrating the same emerging quantities are
different. The functional relations between them provide recursion
relations between the physical parameters, the coupling constants,
which characterize the physics at each scale, and this is precisely
what RG is all about.

Often a major step consists in finding a way of looking at things.
Therefore we stress that the problem we face with the averaged
description in cosmology is effectively a question of how a system
behaves under changes of ``scales". As such it is most naturally
addressed using RG approach, understood here rather as a general
strategy to handle problems of multiple length
scales\footnote{Although the renormalization procedure might seem
purely formal there are important physical ideas behind it, namely,
that of {\it scaling} and {\it universality}.} enabling us to extract
the long distance behaviour of the system by making the scale
successively coarser.  In cosmology, we have the curvature
inhomogeneities and to consistently tackle this problem, we will have
to consider a procedure operating on the metric, not only on or apart
from the matter present.
\bigskip

To provide even more support in favour of the presented above idea,
that RG arguments might be applicable to the problem of the
inhomogeneous universe, let us notice that one can also be guided by
scaling ideas. Scaling is exhibited (approximately or exactly) by many
natural phenomena and mathematical models. The universe, namely, the
distribution of matter in it does exhibit certain scaling properties
\cite{Borgani, Jones}, of which the power law behaviour of the
two-point correlation function for galaxies, clusters and quasars, is
a fair example. Scaling, on its own right, is deeply understood within
the underlying mathematical scaffold which is RG. These are hints
therefore that one can regard the universe as a gravitational
dynamical system not far from criticality (understood intuitively by
analogy with e.g. a ferromagnet).  Later one can also try to qualify
the precise nature of the critical behaviour within the phase
transition context, but this will not be of our concern in the present
paper (apart from a simplified example in section \ref{examplecrit}).
\bigskip

In passing, let us note that interesting results were obtained in
\cite{choptnik} on spherical scalar field collapse and in \cite{abra}
on axisymmetric gravitational waves collapse, which show a surprising
scaling and critical behaviour reminiscent of that found in many
(second-order) phase transition phenomena in condensed matter.
%kam
They could be described, we believe, using RG approach again, treating
the Einstein equations as generating a RG flow on the space of initial
data.
\bigskip

The real space renormalization techniques are mostly applicable to
discretized models, based on a lattice. Therefore we now turn to
describing a suitably discretized manifold model we are going to work
with.

\subsection{Discretized manifold model}

The approach taken is that of a (3+1) formulation of General
Relativity (GR) \cite{ADM}.  Let us suppose we have a differentiable,
compact three-manifold (without a boundary) ${\cal M}$. Generally, in
this case we will always assume that these manifolds possess certain
natural constraints on their diameter and a suitably defined notion of
curvature. The point of this requirement is that the manifolds, or
more precisely the riemannian structures, can then be classified
according to how they can be covered by small metric balls (to be
defined later). Moreover the space of riemannian structures has some
remarkable compactness properties. This is a classical result obtained
by M. Gromov
\cite{Gromov}. On a set of riemannian structures it is possible to
introduce a distance function, the Gromov distance, which roughly
speaking enables one to say something about how close particular
manifolds are to each other.  For the riemannian manifolds which can
be considered close to each other (in the sense of Gromov distance) it
is possible to cover them with the balls arranged in similar packing
configurations \cite{Mauro1}.
\vskip 0.5 cm
In order to define such coverings \cite{GrovePet}, let us parameterize
the geodesics by arc-length, and for any point $p \in {\cal M}$ let
$d_{{\cal M}}(x,p)$ denotes a distance function of the generic point
$x$ from the chosen one $p$. Then for any given $\epsilon >0$ it is
always possible to find an ordered set of points $\{ p_1, \dots ,
p_N\} $ in ${ \cal M} $, so that \cite{GrovePet}
\begin{description}
\item[i)]  the open metric balls (the geodesic balls)
   $ B_{{\cal M}} ( p_i , \epsilon )= \{ x \in {\cal M} |\; d_{{\cal
M}} ( x , p_i) < \epsilon \}, \; i=1, \dots ,N, $ cover ${\cal M}$; in
other words the collection $\{ p_1, \dots, p_N \}$ is an $ \epsilon
$-{\it net} in $ {\cal M}$.
\item[ii)] the open balls $ B_{{\cal M}} ( p_i, \epsilon /2), \; i=1,
\dots ,
  N$ are disjoint, i.e., $\{ p_1, \dots , p_N \}$ is a {\it minimal\/}
$\epsilon$-net in $ {\cal M}$.
\end{description}

It is fair to say that as a consequence of the compactness properties
of the set of riemannian structures that we consider, for each
``length scale $\epsilon$" there exists a finite number of ``model"
geometries which describe, with an $\epsilon-$approximation, any given
riemannian geometry. Namely, given a ball of a certain radius $>
\epsilon$ in any riemannian manifold (with suitable restrictions on
their volume, diameter and sectional curvature, as we have said
earlier) there exists a ball metrically similar (up to an $\epsilon$
scale) in one of the model geometries which does not retain the
details of the original manifold on scales smaller than $\epsilon$.
Roughly speaking, $\epsilon$ is a measure of the typical curvature
inhomogeneity with respect to the model background.  Let us stress
that this is a highly non-trivial result, in the sense that the
metrical properties of the manifolds from an infinite dimensional set
are, up to an $\epsilon$ scale, described by the metrical properties
of just a finite number of model riemannian manifolds.
\bigskip

The $\epsilon-$nets underlying the balls-coverings precisely provide
the discretized manifold model.  This coarse-graining of a manifold
according to Gromov is the most natural coarse-graining one can think
of, pertinent for manifolds with a lower bound to the sectional
curvature. This assumption does not limit the generality of our
analysis which is basically motivated by a concrete physical problem,
whose nature allows us to deal from the beginning with manifolds that
are already in a certain sense quasi-homogeneous (Cf.  comments on the
solvability of the Ricci-Hamilton flow, later on).

In what follows, when speaking of balls we will always mean geodesic
balls here.

\subsection{An empirical averaging procedure}\label{averprocedure}

We assume that we have chosen a particular space-like hypersurface
$\Sigma$ of the four-dimensional manifold, on which the average of a
scalar function $f: \Sigma \rightarrow {\rm I\!R}$ is given as
\begin{equation}\label{aa}
{<f>}_{\Sigma (g)} = \frac{\int_{\Sigma}f d\mu_{g}}{vol(\Sigma,g)} ,
\end{equation}
where $vol(\Sigma,g)=\int_{\Sigma}d\mu_{g}$ and $\mu_{g}$ is the
riemannian measure associated with the three-metric $g$ of $\Sigma$.
Since at this stage we simply wish to put forward a few elementary
geometrical considerations, we do not specify yet the choice of the
hypersurface $\Sigma$ and we do not attribute any particular physical
meaning to the function $f$.
\vskip 0.5 cm
If the geometry of $\Sigma$ is not known on a large scale, we cannot
take (\ref{aa}) as an operational way of defining the average of $f$.
{}From a more pragmatic point of view, supposing that we can only
experience geometry in sufficiently small neighborhoods of a finite
set of instantaneous observers, it makes much more sense to replace
(\ref{aa}) with a suitable average based on the geometrical
information available on the length scale of such observers.\par For
simplicity, given a finite set of instantaneous observers, located at
the points ${x_1,\ldots,x_N}\in\Sigma$, we may assume that these
susceptible to observation regions are suitably small geodesic balls
of radius $\epsilon$, scattered over the hypersurface $\Sigma$ so as
to cover it. In other words, we assume that $\{ x_1,\ldots,x_N\}$ is a
minimal $\epsilon-$net in $\Sigma$.  Further, we denote by
$U_{\epsilon}$ the corresponding set of geodesic balls
$\{B_{\Sigma}(x_{i},\epsilon) \}, i=1,\ldots,N$.  Then we can bound
(\ref{aa}) as

\begin{eqnarray}
<f>_{\epsilon/2}\frac{\sum_hvol(B_h,\epsilon/2)}{vol (\Sigma,g)}
\,\leq\, <f>_{\Sigma(g)}\,\leq\, <f>_{\epsilon}
\frac{\sum_hvol(B_h,\epsilon)}{vol (\Sigma,g)},
\end{eqnarray}

\noindent where

\begin{equation}\label{ab}
{<f>}_{\epsilon}\equiv \frac{\sum_i
\int_{(B_{i},\epsilon)}fd\mu_{g}}{
\sum_i vol (B_{i},\epsilon)} ,
\end{equation}

\noindent and

\begin{equation}
{<f>}_{\epsilon/2}\equiv \frac{\sum_i
\int_{(B_{i},\epsilon/2)}fd\mu_{g}}{
\sum_i vol (B_{i},\epsilon/2)} ,
\end{equation}

\noindent and where we explicitly indicated the dependence
of the averages on a particular ``covering". This suggests to consider
$<f>_{\epsilon}$ as a suitable scale dependent approximation to $<f>$.

\bigskip

There are certain problems lurking that we have to clear up.
Obviously, there are ``unwanted" details affecting the average
function over the discretized manifold, as given by its partition with
a collection of geodesic balls, the immediate one being the underlying
discretization. The important question to ask is what happens to the
average when we change the length scale. Depending on whether we are
actually increasing or decreasing it, respectively less or more
details of the underlying geometry, will be felt by the average
values. The natural philosophy is that over scales big enough no
details should be discerned since the homogeneity and isotropy
prevails. This is the reason why on constant curvature spaces
averaging is well defined since there one can move the balls freely
and deform them, but by so doing no new geometric details that measure
the inhomogeneities will be felt in the averaged values of quantities
we are interested in.

A natural question to ask now is then how the geometry, specifically
curvature inhomogeneities, should depend on scale so that the average
over the balls is scale independent, or equivalently, how do we have
to deform the geometry in order to achieve the scaling limit when size
of the balls matters no more?
\bigskip

First, we give some details on how we calculate the averages according
to (\ref{ab}). In order to do this, we employ a preferred system of
coordinates on the set of balls $\{B_i\}$ given by the local
diffeomorphism
\begin{equation}
\exp_x : T_x {\Sigma} \rightarrow \Sigma
\end{equation}
i.e., we make use of the exponential mapping
\begin{equation}
\varphi_i \equiv \exp {\mid}_{\exp^{-1}B_i \equiv D_i} : D_i
\rightarrow B_i\,,
\end{equation}
where $D_i = D(x_i, \epsilon)$ is the ball in
$T_{x_i}\Sigma$.\footnote{The transition from $(\Sigma, g)$, at a
given moment of time $t_o$, to a tangent space parallels the
prescription in riemannian geometry for measuring \cite{SachsWu}. Let
$U$ be a given neighborhood of an instantaneous observer in $\Sigma$
and suppose it is so small that there is a neighborhood $A$ of $0 \in
T_x \Sigma$ such that $\exp_x:A(\subset T_x\Sigma)\rightarrow
U(\subset \Sigma)$ is a diffeomorphism. One can then replace the
considerations in $(U, g)$ by those in $A$ (with the riemannian
measure pulled back) via $\exp_x^{-1}$. Namely, we can say that an
instantaneous observer in $(\Sigma, g)$ observes the universe with the
help of the exponential mapping, which just means projecting
structures >from an open neighborhood $U\subset \Sigma$ of $x$ by
$\exp^{-1}_x$ and treating them as structures on $T_x\Sigma$.}

On $D_i$ we use polar coordinates and pull-back the riemannian measure
accordingly, namely,
\begin{equation}
{\varphi_i}^{\ast}(\mu_{g})=\theta(t,x_i)dt\otimes dx_i ,
\end{equation}
where $dx_{i}$ denotes the canonical measure (euclidean volume form)
on the unit sphere $D(x_{i},1)=S^2_1 \subset T_{x_{i}}\Sigma$ and
where $dt$ is the Lebesgue measure on ${\rm I\!R}$ ($t \geq 0$).
\bigskip

For $t$ small enough one can prove Puiseux' formula
\begin{equation}\label{puiseux}
\theta(t,x)=t^{n-1} (1-\frac{1}{3}r(x) t^{2} +
{{\cal O}}(t^{2})),
\end{equation}
where $n=dim\,\Sigma$ and $r(x)$ is the Ricci curvature $Ric(g)$ (at
the point $x$).

Using this result we have
\begin{eqnarray}
\int_{B(x_i,t)}fd\mu_g
= \int_{S^2_1}f\theta(t,x)dx_{i}dt.
\end{eqnarray}
\vskip 0.5 cm
Let us consider the asymptotic expansion with respect to $t$
\cite{Lelong}
\begin{equation}
\int_{B(x_i,t)}f d\mu_g = \omega_n t^n [ f(x_i) + \frac{t^2}{2(n+2)}
(\Delta f(x_i)- \frac{R(x_i)}{3}f(x_i))+ {\cal O}(t^2)],
\end{equation}
where $\omega_n$ is the volume of the unit ball of ${\rm I\!R}^n$, $R$
is the scalar curvature at the center of the ball and $\Delta$ the
Laplacian operator relative to the manifold.

Substituting $f=1$ in the above formula we get the asymptotic
expansion of the volume of a geodesic ball:
\begin{equation}
vol(B(x_i,t))=\omega_n t^n( 1- \frac{R(x_i)}{6(n+2)}t^2 + {\cal
O}(t^2)).
\end{equation}

These standard formulae are what we need in order to calculate how the
average value behaves when we change the scale.
\vskip 0.5 cm
Since we are interested in discussing how $<f>_{\epsilon}$ behaves
upon changing the radius of the balls $\{B(x_i,\epsilon)\}$, let us
consider the average $<f>_{{\epsilon}_o+\eta}$, with $\eta$ a positive
number with $\eta/\epsilon_o\ll 1$. According to the formulae recalled
above, we can write

\begin{equation}
<f>_{{\epsilon}_o+\eta}= \frac{\sum_i [f_i+(\frac{\Delta f_i-R_i
f_i/3}{2(n+2)} )(\epsilon_o+\eta)^2]}{\sum_i [1-
\frac{R_i}{6(n+2)}(\epsilon_o+\eta)^2]}+{\cal O}((\epsilon_o+\eta)^4),
\end{equation}
where we have introduced somewhat simplified but otherwise obvious
notation.  Upon expanding this expression in $\eta$, we get to leading
order (${\cal O}(\epsilon_o^4)$ in $\epsilon_o$, and ${\cal
O}([\frac{\eta}{\epsilon_o}]^2)$ in $\frac{\eta}{\epsilon_o}$),

\begin{equation}
<f>_{\epsilon_o+\eta}\simeq <f>_{\epsilon_o} +
\frac{1}{n+2}{<\Delta
f>}_{\epsilon_o}{\epsilon_o}^2\frac{\eta}{\epsilon_o} +
\frac{1}{3(n+2)}[<R>_{\epsilon_o}<f>_{\epsilon_o}-
<Rf>_{\epsilon_o}]{\epsilon_o}^2\frac{\eta}{\epsilon_o},
\end{equation}

\noindent where $<f>_{\epsilon_o}$
is the average of the function $f$ over the set of $N$ instantaneous
observers $U_{\epsilon_o}$, (with similar expressions for
$<R>_{\epsilon_o}$, $<Rf>_{\epsilon_o}$ and
$<{\Delta}f>_{\epsilon_o}$). Thus, under a change of the cutoff we can
write

\begin{equation}\label{rg}
{\epsilon_o}
\frac{d}{d\eta}{<f>}_{\epsilon_o+\eta}|_{\eta/\epsilon_o=0}=
\frac{1}{n+2}{<\Delta f>}_{\epsilon_o}+
\frac{1}{3(n+2)}[<R>_{\epsilon_o}<f>_{\epsilon_o}-<Rf>_{\epsilon_o}]
\end{equation}
to leading order.  In the next sections we will discuss the
consequences of these formulae and the connection of our averaging
procedure with the Ricci-Hamilton flow.
\bigskip

\section{The Renormalization Group view}

\subsection{Block variables and recursion relations}

The real space RG technique is based on the recursive introduction of
{\em block variables}. A method of ``blocking" is trivial to introduce
on regular lattices. In particular, in the case of the Ising model
($2-$dim) it consists of a subdivision of the spin system into cells,
which supposedly interact in a similar way as the original spins.
This can be done by introducing the block-spin variables via the
majority rule or decimation procedure \cite{Binney}. The behaviour of
blocked lattice on large scales is equivalent to the behaviour of the
original lattice corresponding to a different temperature. The slopes
of the parameters' surface close to the {\em critical fixed point}
determine the macroscopic characteristics of the model (see e.g.
\cite{Binney}).
\vskip 0.5 cm
In the context of the problem we are considering, the application of
RG method forces us to invent some analogue of Kadanoff's blocking
(block-spin transformation) applied to the geometry itself. This
appears to be a difficult problem since in a general case when the
geometry is curved the ``lattice" itself takes on a dynamical
r\^{o}le. Moreover, since the manifold $\Sigma$ we are dealing with is
compact, (closed, without boundary), we must implement a
Renormalization Group strategy in a finite geometry, and thus the
relevant phenomena are here related to {\em finite size scaling}.
Roughly speaking, the size of our manifold is characterized by a
length scale, say $L$, (actually the volume, the diameter and possibly
bounds on curvatures), which is large in terms of the microscopic
scale (the radius of the typical geodesic ball coverings we shall use
in the blocking procedure). A continuous theory, describing the
(universal) properties of the field $f$ on $\Sigma$, arises when the
correlation length associated with the distribution of $f$ is large
(of the order of $L$, or bigger). This being the underlying rationale,
let us proceed and be guided by the formula (\ref{rg}).

\vskip 0.5 cm
Let us consider our system as nearly infinite, i.e. the manifold
$\Sigma$ divided by the collection of the geodesic balls of radius
$\epsilon_m\equiv (m+1)\epsilon_o$ for $m=0,1,2,\ldots$, and where
$\epsilon_o$, the chosen cutoff, is much smaller than the typical
length scale associated with $\Sigma$, (this length scale can be
identified with the injectivity radius of the manifold). Each ball
will be labelled by $k$ in the sequel and the original geodesic
packing-covering is for $m=0$.  We can introduce this way a convenient
notation for the integral of the generic function $f$ over the ball
$B(x_k,{\epsilon}_m)$ as
\begin{equation}
\psi_m(k;f)\equiv
\int_{B(x_k,{\epsilon}_m)}fd{\mu}_g,
\end{equation}
which can be seen as the block variable since it allows us to
eliminate >from the distribution of the field $f$ all fluctuations on
scales smaller than the cutoff distance $\epsilon_m$. We wish to
emphasize that if the geometry of the ball $B(x_k,{\epsilon}_m)$ is
not flat, then the definition of ${\psi}_m(k;f)$ can be interpreted as
that of a {\em weighted} sum over a flat ball, namely,

\begin{equation}
\psi_m(k;f)\equiv
\int_{{\exp}^{-1}B(\epsilon_m)}f
{\theta}(t,x)dt\otimes dx\,,
\end{equation}
 where the weight ${\theta}(t,x)$ is provided by Puiseux' formula
(\ref{puiseux}).
\vskip 0.5 cm

If we consider the covering of $B(x_k,{\epsilon}_{m+1})$ induced by
the geodesic balls $\{B(x_j,\epsilon_m)\}$, i.e., the collection of
$N_k(m,m+1)$ open sets $\{B(x_k,\epsilon_{m+1})\} \cap
\{B(x_j,\epsilon_m)\}$, then the above definition of block variables
can be written recursively in terms of the values the function $f$
takes correspondingly to balls of larger and larger radii. To this
end, let us consider a partition of unity
$\{{\xi}_h\}_{h=1,\ldots,N_k(0,m+1)}$, subordinated to the covering of
the generic enlarged ball $B(x_k,{\epsilon}_{m+1})$, induced by the
geodesic balls $\{B(x_h,\epsilon_o)\}$. Namely, a set of smooth
functions such that: $0\leq \xi_h\leq 1$, for each $h$; the support of
each $\xi_h$ is contained in the corresponding $B(x_h,\epsilon_o)$;
and $\sum_h\xi_h(p)=1$, for all $p\in B(x_k,\epsilon_{m+1})$. \par
\vskip 0.5 cm
Under such assumptions, the block variables $\psi_m(k;f)$ can be
written recursively as

\begin{eqnarray}\label{recursive}
\psi_o(k;f)&= &
\int_{B(x_k,{\epsilon}_o)}\xi_kfd{\mu}_g, \nonumber\\
\psi_{m+1}(h;f)&= &
\sum_k^{N_h(m,m+1)}\psi_m(k;f).
\end{eqnarray}
\vskip 0.5 cm
Indeed, we have

\begin{eqnarray}
\psi_1(h;f)&= &
\int_{B(y_h,{\epsilon}_1)}fd{\mu}_g=\sum_k^{N_h(0,1)}
\int_{B(x_k,{\epsilon}_o)}\xi_kfd{\mu}_g\nonumber\\
&= & \sum_k^{N_h(0,1)}\psi_o(k;f),
\end{eqnarray}

\noindent and

\begin{eqnarray}
\psi_2(j;f)&= &
\int_{B(z_j,{\epsilon}_2)}fd{\mu}_g=\sum_h^{N_j(1,2)}
\int_{B(y_h,{\epsilon}_1)}\xi_hfd{\mu}_g\nonumber\\
&= &\sum_h^{N_j(1,2)}\sum_k^{N_h(0,1)}
\int_{B(x_k,{\epsilon}_o)}\xi_k\xi_h fd{\mu}_g\nonumber\\ &= &
\sum_h^{N_h(1,2)}\psi_1(h;f),
\end{eqnarray}
where we have exploited the fact that the functions $\{\xi_k\xi_i\}$
have $\xi_k\xi_i(p)=0$, except for a finite number of indices $(k,i)$,
and $\sum_k\sum_i\xi_k\xi_i(p)=1$, for all $p\in \Sigma$. The above
expressions readily generalize for every $m$.

\vskip 0.5 cm
More explicitly, we can write these block variables as
\vskip 0.5 cm

\begin{eqnarray}
& &\psi_{m+1}(k;f)=
\int_{B(x_k,{\epsilon}_{m+1} )}fd{\mu}_g=\nonumber\\
& &{\omega}_n \epsilon_m^n\sum_h^{N_k(m,m+1)}\left[f(h)\xi_h+
\left(\frac{\Delta(f\xi_h)(h)-R(h)f(h)\xi_h/3}
{2(n+2)} \right)\epsilon_m^2 +{\cal O}(\epsilon_m^4)\right].
\end{eqnarray}
\vskip 0.5 cm

Notice that in terms of the block variables $\psi_m(k;f)$ we can
rewrite the empirical averaging (\ref{ab}) as
\be
\label{empirical}
<f>_{\epsilon_m}(\psi_m)=\frac{\sum_k^{N(\epsilon_m)}
\psi_m(k;f)}{\sum_k^{N(\epsilon_m)}vol[B(x_k,\epsilon_m)]},
\ee
where $N(\epsilon_m)$ denotes the number of distinct
$B(x_k,\epsilon_m)$ balls providing a minimal $\epsilon_m$-covering of
the manifold $\Sigma$. Thus, when $m$ is sufficiently large, the
variation in $<f>_{\epsilon_m}(\psi_m)$ under a block transformation
$\psi_m(k;f)\to\psi_{m+1}(h;f)$ is given (to leading order) by

\begin{eqnarray}
& &<f>_{\epsilon_{m+1}}(\psi_{m+1}) -
<f>_{\epsilon_{m}}(\psi_{m})\simeq\\ &
&\frac{1}{n+2}{<\Delta(f\xi_h)>}_{\epsilon_m}{\epsilon_m}^2\frac{1}{m+1}
+
\frac{1}{3(n+2)}[<R>_{\epsilon_m}<f\xi_h>_{\epsilon_m}-
<Rf\xi_h>_{\epsilon_m}]{\epsilon_m}^2\frac{1}{m+1}.\nonumber
\end{eqnarray}
\vskip 0.5 cm

The above choice of block variables brings out the coupling between
averaging a scalar field over a manifold and the presence of
fluctuations in the curvature of the underlying geometry.
\vskip 0.5 cm
In order to be more precise, let us assume that the variables $f(k)$
are randomly distributed according to some probability law
$P(\{f(k)\})$ (later on we shall come back to this point with a
definite prescription).  Upon blocking the system and thus
renormalizing the variables $f(k)$ by increasing the scale size, the
probability distribution $P(\{f(k)\})$ induces a corresponding
probability distribution on the variables $\psi_m$, {\it viz.},
$P(\{\psi_m(k;f)\})$.
\vskip 0.5 cm
 From equation (\ref{recursive}) it is clear that if the geometrical
properties of any two balls, $B(x_i,\epsilon_{{\bar m}})$ and
$B(x_j,\epsilon_{{\bar m}})$ (with $B(x_i,\epsilon_{{\bar m}})\cap
B(x_j,\epsilon_{{\bar m}})=\emptyset$), are not correlated then the
corresponding block variables, $\psi_{{\bar m}}(i;f)$ and $\psi_{{\bar
m}}(j;f)$, are uncorrelated. Such length scale $L\equiv\epsilon_{{\bar
m}}$ characterizes the correlation (or persistence) length of the
manifold $(\Sigma,g)$. It is a measure of the typical linear dimension
of the largest ball exhibiting a correlated spatial structures.  This
correlation length can be seen in close analogy with the usual
correlation length (usually denoted by $\xi$) in condensed matter
systems.  It depends there upon the coupling constants in particular
upon temperature, and diverges to infinity at the phase transition
point.
\vskip 0.5 cm
Since $g$ plays here the r\^{o}le of a running coupling, or if you
prefer, of ``temperature", the existence of a finite correlation
length corresponds to a rather ``irregular", crumpled geometry (as
seen on scales of the order of $L$), or, equivalently, a {\it high
temperature phase} of our system.
\vskip 0.5 cm
According to the central limit theorem it follows, for $m$ large
enough, ($\epsilon_m\gg L$), that the block variables $\psi_m(k;f)$,
being the sum of uncorrelated random variables, are normally
distributed, (let us say around zero, for simplicity), with a variance
\begin{eqnarray}
E_P(\psi_m^2(i;f))=N^{(\epsilon_o,m)}{\bar \chi},
\end{eqnarray}
where $E_P(\ldots)$ denotes the expectation according to the
probability law $P(\{\psi_m\})$, ${\bar \chi}$ is related to the
variance of the variables $f(k)$, $R(k)$, and $N^{(\epsilon_o,m)}$
denotes the number of $\epsilon_o$-balls in the $m$-ball
$B(x_i,\epsilon_m)$.
\vskip 0.5 cm
Thus, irrespective of the details of the local distribution of the
random variables $f(k)$ and $R(k)$, we can write for the distribution
of $\{\psi_m(k;f)\}_1^{N}$ over $(\Sigma,g)$

\begin{eqnarray}
dP(\{\psi_m\})=\prod_k\left[ d\psi_m(k;f){\left( 2\pi
N^{(\epsilon_o,m)}{\bar \chi} \right)}^{-1/2}
\exp\left[-\psi_m^2(k;f)/2N^{(\epsilon_o,m)}{\bar \chi}     \right]
\right].
\end{eqnarray}

This shows that by rescaling the block variables $\psi_m(k;f)$
according to

\begin{eqnarray}
\phi_m(k;f)\equiv {\left[N^{(\epsilon_o,m)} \right]}^{-1/2}
\psi_m(k;f),
\end{eqnarray}
we get new block variables with a finite variance as $m\to\infty$, and
for {\it random} metrics $(\Sigma,g)$ we can write
\begin{eqnarray}
dP(\{\phi_m\})=\prod_k\left[ d\phi_m(k;f){\left( 2\pi{\bar \chi}
\right)}^{-1/2}
\exp\left[-\phi_m^2(k;f)/2{\bar \chi} \right] \right].
\end{eqnarray}
\vskip 0.5 cm

The above remarks, paradigmatic of the real space Renormalization
Group philosophy, show that the definition of a sensible blocking
procedure, in our geometrical setting, consists of a transformation
increasing the scale size, realized by passing from the variables
$f(k)$ to the variables $\psi_m(k;f)$, (namely, by taking the average
over all values of $f$ in a larger and larger ball), followed by a
rescaling obtained by dividing $\psi_m(k;f)$ by a suitable power of
the number, $N^{(\epsilon_o,m)}$, of elementary $\epsilon_o$-balls
contained in the $\epsilon_m$-ball considered, (for random geometries
this power is $1/2$).
\vskip 0.5 cm
Following standard usage, and in order to arrive at an interesting
geometrical notion of blocking, we assume that for a generic metric
$g$ this rescaling follows by dividing $\psi_m(k;f)$ by ${\left[
N^{(\epsilon_o,m)}
\right]}^{\omega_m}$, where $\omega_m$ will in general depend on $m$.
Thus, the rescaled blocked variables of relevance are

\begin{eqnarray}
\phi_m(k;f)={\left[ N^{(\epsilon_o,m)} \right]}^{-\omega_m}
\psi_m(k;f).
\end{eqnarray}

The value of $\omega_m$ will be fixed by the requirement that, as
$m\to\infty$, and for some (critical) metric $g_{crit}$, (in general
for an open set of such metrics), such normalized large scale block
variables have a limiting probability distribution with a finite
variance. Namely,
\begin{eqnarray}
\label{normalization}
\lim_{m\to\infty}P_m(\{\phi_m(k;f)\})&= &
P_{\infty}(\{\phi_{\infty}(f)\}), \nonumber\\
E_{P_{\infty}}(\phi_{\infty}^2(f))=1.
\end{eqnarray}
\vskip 0.5 cm

Notice that if $(\Sigma,g)$ is a {\it nice} manifold, e.g. a constant
curvature simply connected three-manifold, we are obviously expecting
that the corresponding $\omega_m$ is

\begin{eqnarray}
\omega_m(\Sigma,g)=1.
\end{eqnarray}

In general, we can assume that there is a set of critical metrics and
a corresponding $\omega$, such that the above requirements
(\ref{normalization}) are satisfied. Such critical metrics are not
necessarily constant curvature metrics and the corresponding $\omega$
is not necessarily $1$, (one may conjecture that $1/2\leq\omega\leq
1$, as happens in the Renormalization Group analysis of many magnetic
systems). At this stage, this is only a tentative assumption in order
to arrive at an interesting concept of geometric Renormalization Group
in our setting. Later on, we shall see that such assumptions are
justified by exhibiting examples of such non-trivial metrics.

\bigskip
\section{Averaging  matter and geometry}

Until now, our discussion has addressed mainly geometrical issues and
the function $f$ entering the cut-off dependent averaging
$<f>_{\epsilon}$ was not specified.  Now we wish to apply the obtained
results to the averaging of the matter sources, namely, the matter
density $\rho$, the spatial stress tensor $s_{ab}$, and the momentum
density $J_a$, entering in the phenomenological description of the
matter energy-momentum tensor with respect to the instantaneous
observers comoving with $\Sigma$, {\it viz.}
\begin{equation}
T_{ab}=\rho u_a u_b+J_a u_b+J_b u_a+s_{ab},
\end{equation}
where ${\bf u}$ is a unit, future directed normal to the slice
$\Sigma$, ($u^au_a=-1$).\par
\vskip 0.5 cm
In fact, since in the present epoch the universe is mainly matter
dominated, i.e. the pressure can be safely neglected, it would in
principle be justified to start our analysis with
\begin{equation}
T_{ab}=\rho u_a u_b+J_a u_b+J_b u_a,
\end{equation}
for the stress-energy tensor. Even the terms involving the momentum
density can be eliminated if we pick up a sensible slicing $\Sigma_t$
(the comoving frame) provided that the cosmological matter fluid is an
irrotational fluid in equilibrium. In any case, wishing to maintain
the following discussion to a sufficient degree of generality, without
particular restrictive assumptions on the matter sources, we assume in
our analysis that the matter energy-momentum tensor has the perfect
fluid form
\begin{equation}
T_{ab}=\rho u_a u_b+J_a u_b+J_b u_a+g_{ab}p,
\end{equation}
with a pressure $p$ which is {\it a priori} not-vanishing, and where
${g}_{ab}$ is the three-metric of $\Sigma$, namely, ${g}_{ab}=
g^{(4)}_{ab}+n_a n_b$, $g^{(4)}_{ab}$ denoting the space-time
metric\footnote{The fact that we are adopting a stress tensor with a
non-vanishing energy flux can be traced back to the observation that
even in an exact $k=0$ or $k=+1$ FLRW universe there can be a non-zero
particle flux tilted with respect to the frame of the fundamental
observers, even though $T_{ab}$ has the perfect fluid form. This is
possible if the macroscopic quantities (e.g. a particle flux) are
derived from kinetic theory with anisotropic distribution function
$f(x^i, p^a)$ \cite{EMT}.}.
\vskip 0.5 cm
As argued in the previous section, smoothing-out the matter sources as
described by a set of instantaneous observers (represented by the
three-dimensional hypersurface $\Sigma$), means eliminating from the
distribution of such sources on $\Sigma$ all fluctuations on scales
smaller than the cutoff distance $\epsilon$, leaving an effective
probability distribution of fluctuations for the remaining degrees of
freedom. The underlying philosophy being that this effective
distribution has the same properties as the original one at distances
much larger than $\epsilon$ (i.e. for fluctuations with wavelengths
much larger than $\epsilon$).\par
\vskip 0.5 cm
In order to implement this idea along the geometrical lines discussed
in the last section, we first need to better specify what we mean by
the assignment of a collection of instantaneous observers (endowed
with clocks) on $\Sigma$. Since we are adopting a Hamiltonian point of
view, such observers are specified by the assignment, on the
(abstract) three-manifold $\Sigma$, of the lapse function $\alpha$ and
the shift vector field ${\bf
\alpha}^i$.
The former provides the local rate of the coordinate clocks of such
observers, while the latter is the three-velocity vector of the
observers with respect to the set of instantaneous observers at rest
on $\Sigma$.\par
\vskip 0.5 cm
The macroscopic variables of interest characterizing the matter
sources are in the present framework, the matter density $\rho$, and
the momentum density ${\bf J}$. Actually, it would make more sense to
consider the cosmological fluid phenomenologically, described by a
pressure $p$, a baryon number density $n_b$, energy density $\rho$,
specific entropy $s$, and temperature $T$; such variables being
related by
\begin{eqnarray}
dp=n_bdh-n_bTds
\end{eqnarray}

\noindent and

\begin{eqnarray}
h=\frac{p+\rho}{n_b},
\end{eqnarray}

\noindent where $h$ is the specific enthalpy. However, for simplicity,
we shall in the sequel consider a barotropic fluid.\par
\vskip 0.5 cm

Thus, the field $f$ characterizing the matter sources, as described by
the instantaneous observers on $\Sigma$, is given by

\begin{equation}
\label{sources}
f\equiv \alpha{\rho} + \alpha^iJ_i.
\end{equation}
Notice that $2{\alpha}p(\rho)$ is the hamiltonian density of the fluid
in the Hamiltonian formulation of Taub's variational principle for
relativistic perfect fluids \cite{Moncrief}; we also assume that the
{\it dominant energy condition} holds.

The averaging of the matter sources along the lines described in the
previous section would then require considering a finite set of
instantaneous observers, $\{x_1,\ldots,x_N\}$, located on $\Sigma$ and
setting a standard for the cutoff distance $\epsilon_o$ over which the
(experimental) distribution of matter sources, (i.e. the probability
that the matter variables, ${\rho}(i)$ and ${\bf J}(i)$, conform to a
given distribution ${\rho}(i)d\rho$, ${\bf J}(i)d{\bf J}$), is
determined.  Then we proceed with the blocking prescription for
eliminating unwanted degrees of freedom and consider the behaviour as
the averaging regions become larger and larger.\par
\vskip 0.5 cm
We may decide to treat the riemannian geometry of $\Sigma$ as
uncoupled with the matter sources if we are simply interested in
smoothing-out the sources, or if we wish to consider fluctuations of
the matter as essentially uncoupled to the fluctuations in geometry.
Namely, the curvature fluctuations appearing in the definitions of the
block variables, $\psi_m(k;f)$ and $\phi_m(k;f)$, can be thought of,
under appropriate circumstances, as independent random variables with
a given distribution. \par

In general, however, the sources are coupled to the gravitational
field, and we ought to treat the full dynamical system, the
cosmological fluid plus the geometry of $\Sigma$, in the procedure of
blocking.  Moreover, we should bear in mind that without taking
explicitly into account the backreaction of the geometry one cannot
really provide a reasonable averaging procedure for the sources.
\vskip 0.5 cm
In order to do so, we consider for a given cutoff distance
$\epsilon_o$, the variables ${\rho}(k)$ and ${\bf J}(k)$ associated
with a minimal geodesic ball covering $\{B(x_k,\epsilon_o)\}_{k=1}^N$.
The change of the cutoff is naturally realized by considering balls
$\{B(x_l,\epsilon_m=(m+1)\epsilon_o)\}$ with $m=0,1,\ldots$. The block
variables, $\psi_o(k;\rho,{\bf J})$ and $\psi_m(k;\rho,{\bf J})$, are
defined according to (\ref{recursive}) with $f$ given by
(\ref{sources}).  This transformation can be seen as thinning-out the
degrees of freedom which is at the heart of any coarse-graining.\par

The original cutoff $\epsilon_o$ is chosen to set the scale over which
General Relativity is experimentally verified and it can be taken as,
say, the scale of planetary systems. We can thus safely assume that
the Einstein field equations hold on that scale. It is however rather
impossible to provide a mathematical model of the distribution of
matter in the universe going down to such fine scales; besides this
task would be impractical.  What one does instead is to use continuous
functions assuming that they represent appropriately ``volume
averages". The results of such an averaging in an inhomogeneous medium
obviously depend on the scale.  The point is that if the Einstein
equations hold on the scale where they have been verified (here taken
to be that of the planetary scale), then they do not seem to hold {\it
a priori} on larger, cosmological, scales that require averaging. To
see this, notice that the Einstein tensor $\tilde{G}_{\mu\nu}$,
calculated from an ``averaged" metric $\bar{g}_{\mu\nu}$, cannot be
equal to the Einstein tensor $\bar{G}_{\mu\nu}$ which was first
calculated >from the fine-scale metric $g_{\mu\nu}$ and then averaged.
This is so due to the non-commutativity of ``averaging the metric''
with calculating the Einstein tensor being strongly non-linear in the
metric components.
\bigskip

Below we shall assume a Hamiltonian point of view. Then the
probability $P(\{\rho(k),{\bf J}(k)\})$ that the matter variables,
according to the records of the instantaneous observers
$\{B(x_k,\epsilon);\alpha(x_k),{\alpha}^i(x_k)\}$ in $\Sigma$, take on
some particular set of values, $\{\bar{\rho}(k), \bar{{\bf J}}(k)\}$,
is given by the equation
\begin{eqnarray}
P(\{\psi_o(k;\rho,{\bf J})\}_{1,\ldots,N})=\frac{1}{Z}\exp[-{
H}(\{\rho(k),{\bf J}(k)\})],
\end{eqnarray}
where $Z$ is a normalization factor, and ${ H}(\{\rho(k),J(k)\})$ is
the Hamiltonian associated with the matter variables $\rho$ and ${\bf
J}$.  Namely,
\begin{eqnarray}
P(\{\psi_o(k;\rho,{\bf J})\}_{1,\ldots,N})=
\frac{\exp\left[ -<\alpha\rho+\alpha^hJ_h>_{\epsilon_o}
\right]}
{\sum_{\rho,J}\exp\left[ -<\alpha\rho +\alpha^hJ_h>_{\epsilon_o}
\right]},
\end{eqnarray}
where $\sum_{\rho,J}\ldots$ denotes summation over all possible
configurations of the matter variables, $\rho$ and ${\bf J}$, that can
be experienced by the instantaneous observers
$\{B(x_k,\epsilon);\alpha(x_k),{\alpha}^i(x_k)\}$ in $\Sigma$.

\vskip 0.5 cm
The validity of General Relativity at this scale implies that this
Hamiltonian is a part of ${\cal H}_{ADM}(\{\rho(k),{\bf J}(k)\})$, the
Arnowitt-Deser-Misner Hamiltonian, associated with the data of the
three-geometry $g_{ab}$ of $\Sigma$, its conjugate momentum
${\pi}_{ab}$ and of the matter variables $\rho$ and ${\bf J}$,
evaluated in correspondence with the
$\{B(x_k,\epsilon)\}$-approximation associated with the net of points
$\{x_k\}$.  Namely,
\begin{eqnarray}
{\cal H}_{ADM}(\{\rho(k),{\bf J}(k)\})|_{\epsilon}=<\alpha {\cal
H}(g,\pi, G,\rho)+ {\alpha}^h{\cal H}_h(g,\pi,G,{\bf J})>_{\epsilon}
\sum_i vol(B(x_i,\epsilon)),
\end{eqnarray}
where
\begin{eqnarray}
{\cal H}(g,\pi, G,\rho)=(det(g))^{-1/2}[\pi^{ab}\pi_{ab}-
\frac{1}{2}(\pi^a_a)^2]-(det(g))^{1/2}R(g)+8\pi(det(g))^{1/2}G\rho,
\end{eqnarray}
\noindent and
\begin{eqnarray}
{\cal H}_a(g,\pi,G,{\bf J})=-2{\pi_a^b}_{;b}+16\pi(det(g))^{1/2}GJ_a,
\end{eqnarray}
\noindent and where the momentum  conjugate to the three-metric
$g_{ab}$ is given in terms of the second fundamental form $K_{ab}$ of
the embedding of $\Sigma$ in the resulting space-time, as
\begin{eqnarray}
\pi^{ab}=(det(g))^{1/2}(K^{ab}-K^c_cg^{ab}).
\end{eqnarray}
\vskip 0.5 cm

We wish to recall that the lapse and the shift appear in the
Hamiltonian as arbitrary Lagrange multipliers (their evolution is not
specified by the equations of motion), and as such they enforce the
constraints
\begin{eqnarray}
{\cal H}(g,\pi, G,\rho)=(det(g))^{-1/2}[\pi^{ab}\pi_{ab}-
\frac{1}{2}(\pi^a_a)^2]-(det(g))^{1/2}R(g)+8\pi(det(g))^{1/2}G\rho=0
\end{eqnarray}
\noindent and
\begin{eqnarray}
{\cal H}_a(g,\pi,G,{\bf J})=-
2{\pi_a^b}_{;b}+16\pi(det(g))^{1/2}GJ_a=0.
\end{eqnarray}
\vskip 1 cm
As is well known, these constraints are related to the invariance of
the theory under the (four-dimensional) diffeomorphism group of the
space-time resulting from the evolution of the initial data satisfying
them. The momentum constraint, ${\cal H}_a(g,\pi,G,{\bf J})=0$,
generates the (spatial) diffeomorphisms into $\Sigma$, while the
Hamiltonian constraint, ${\cal H}(g,\pi, G,\rho)=0$, generates the
deformation of the manifold $\Sigma$ in the resulting space-time, i.e.
the dynamics. (Notice that such deformations can be interpreted as
four-dimensional diffeomorphisms only after that space-time has been
actually constructed).

The Hamiltonian ${\cal H}_{ADM}(\{\rho(k),{\bf J}(k)\})$, apart from
the lapse and the shift, depends on the three-metric $g_{ab}$ on
$\Sigma$ and its conjugate momentum $\pi_{ab}$, the gravitational
coupling $G$, and on the configuration which the matter variables,
$\rho(k)$ and ${\bf J}(k)$, take on the set of instantaneous
observers, $\{B(x_k,\epsilon);\alpha(x_k),{\alpha}^i(x_k)\}$, chosen
to describe the distribution of matter at the given length scale
$\epsilon$.\par
\vskip 0.5 cm
The basic question to understand is how the block transformations,
\begin{eqnarray}
\{\phi_{m}(k;\rho,{\bf J})\}
\to \{\phi_{m+1}(k;\rho,{\bf J})\},
\end{eqnarray}
followed by rescaling, affect the Hamiltonian associated with the
matter variables $\rho$ and ${\bf J}$, and then discuss how this in
turn affects the full Hamiltonian ${\cal H}_{ADM}$.
\vskip 0.5 cm

Starting from the probability distribution $P(\{\psi_o(k;\rho,{\bf
J})\}_{1,\ldots,N})$ we can inductively define the probability
distribution $P^{(m+1)}(\{\psi_{m+1}(k;\rho,{\bf J})\})$ of the block
variables $\{\psi_{m+1}(k;\rho,{\bf J})\}$ (and of the corresponding
rescaled variables $\{\phi_{m+1}(k;\rho,{\bf J})\}$). Since the block
variables $\{\psi_{m+1}(k;\rho,{\bf J})\}$ are recursively obtained
from the knowledge of the block variables at the $m^{th}$ stage,
$\{\psi_{m}(k;\rho,{\bf J})\}$, the distribution
$P^{(m+1)}(\{\psi_{m+1}(k;\rho,{\bf J})\})$ only depends on the
knowledge of $P^{(m)}(\{\psi_{m}(k;\rho,{\bf J})\})$. We can formally
write
\begin{eqnarray}
P^{(m+1)}(\{\psi_{m+1}(k;\rho,{\bf J})\})=\sum_{\{\psi_m(k;\rho,{\bf
J})\}} P^{(m)}(\{\psi_m(k;\rho,{\bf J})\}),
\end{eqnarray}
where the sum is over the probabilities of all the configurations
$\{\psi_m(k;\rho,{\bf J})\}$ consistent with the configuration
$\{\psi_{m+1}(k;\rho,{\bf J})\}$ of the block variables.
\vskip 0.5 cm

As usual, this allows us to define the effective Hamiltonian for the
matter variables after $m$ iterations of the block transformation,
according to
\be
\label{effective}
P^{(m)}(\{\psi_m(k;\rho,{\bf J})\})\equiv\frac{1}{Z^{(m)}}\exp[-{
H}^{(m)}(\{\psi_m(k;\rho,{\bf J})\})],
\ee
\noindent where
\begin{eqnarray}
Z^{(m)}\equiv\sum_{\{\psi_m(k;\rho,{\bf J})\}}
\exp[-{H}^{(m)}(\{\psi_m(k;\rho,{\bf J})\})].
\end{eqnarray}
\vskip 0.5 cm
Such ${ H}^{(m)}(\{\psi_m(k;\rho,{\bf J})\})$ are defined up to an
additive constant term (e.g. \cite{Binney}). If we also stipulate, as
is standard usage in the Renormalization Group approach, that the
effective Hamiltonian ${H}^{(m+1)}(\{\psi_{m+1}(k;\rho,{\bf J})\})$
takes the same functional form as ${H}^{(m)}(\{\psi_m(k;\rho,{\bf
J})\})$, i.e., if
\begin{eqnarray}
P^{m+1}(\{\psi_{m+1}(k;\rho,{\bf J})\}_{1,\ldots,N})=
\frac{\exp\left[ -
<\alpha\rho^{(m+1)}+\alpha^hJ_h^{(m+1)}>_{\epsilon_{m+1}}
\right]}
{\sum_{\rho,J}\exp\left[ -<\alpha\rho^{(m+1)} +
\alpha^hJ_h^{(m+1)}>_{\epsilon_{m+1}}\right]},
\end{eqnarray}
then this indeterminacy can be transferred to the {\it effective}
matter variables $\rho^{(m+1)}$ and ${\bf J}^{(m+1)}$, in terms of
which ${H}^{(m+1)}(\{\psi_{m+1}(k;\rho,{\bf J})\})$ is defined.
\vskip 0.5 cm
It is immediately checked that such effective matter variables are
defined by (\ref{effective}) up to the transformations
\begin{eqnarray}
\label{indeterminacy}
\rho^{(m+1)} &\to & \rho^{(m+1)}+h,\nonumber\\
{\bf J}^{(m+1)} &\to & {\bf J}^{(m+1)} + {\bf v},
\end{eqnarray}
where $h$ and ${\bf v}$ are, respectively, a scalar function
(sufficiently regular) and a vector field defined on $\Sigma$.
\vskip 0.5 cm

 Among the various normalization conditions that we may adopt, in
order to avoid the indeterminacy connected with (\ref{indeterminacy}),
the natural one comes about {\it by requiring the divergence and the
Hamiltonian constraints to hold at each stage of the renormalization}.
This being the case, upon coupling matter to the geometry, the lapse
$\alpha$ and the shift $\alpha^i$ maintain their r\^{o}le of Lagrange
multipliers enforcing the (four-dimensional) diffeomorphism invariance
of the theory.\par

\vskip 0.5 cm
In this way, the renormalization is set up so that the constraints, if
initially satisfied, will remain satisfied at every renormalization
step.  The three-metric $g$, (and the second fundamental form), in its
r\^{o}le of {\it running coupling} describing the interaction between
matter and geometry, is then governed through (Coulomb like) effects
that are expressed by these constraint equations. In this sense, the
Coulomb like part of the gravitational interaction is driving the
renormalization mechanism, and such requirements imply that the full
effective Hamiltonian, (matter plus geometry), takes on the standard
ADM form pertaining to gravity interacting with a barotropic fluid at
every stage of the renormalization.
\vskip 0.5 cm
According to the remarks above, we can consider as an independent
parameter in the (effective) fluid Hamiltonian ${
H}^{(m)}(\{\psi_m(k;\rho,{\bf J})\})$ the three-metric $g^{(m)}_{ab}$
of the three-manifold $\Sigma$, whereas the matter density
$\rho^{(m)}$ and the current density ${\bf J}^{(m)}$ are at each stage
connected to $g^{(m)}_{ab}$ and $K^{(m)}_{ab}$ by the Hamiltonian and
divergence constraints that hold at each stage. Then the effect of the
renormalization induced by the blocking procedure
$\{\psi_m(k;\rho,{\bf J})\}\to\{\psi_{m+1}(k;\rho,{\bf J})\}$ and the
corresponding rescaling, can be symbolized as a non-linear operation
acting on the metric $(g^{(m)}_{ab})$ so as to produce the metric
$(g^{(m+1)}_{ab})$, i.e.
\be
\label{RGmapping}
(g^{(m+1)}_{ab})={\cal R}(g^{(m)}_{ab}),
\ee
whereas the renormalization of the second fundamental form $K_{ab}$ is
generated by the {\it linearization} of (\ref{RGmapping}), as will be
discussed in section 5.

This way, this renormalization transformation ${\cal R}$ defines a
trajectory in the space of riemannian metrics of $\Sigma$. One moves
on such trajectory by discrete jumps. However, in what follows we
shall replace such discrete dynamical system by a smooth dynamical
system, describing renormalizations of the parameters $(g_{ab})$ and
$(K_{ab})$.
\vskip 0.5 cm
The deformation of the initial data set for the Einstein field
equations, symbolically denoted by ${\cal R}$ above in
(\ref{RGmapping}), realizes in fact a {\it formal}, (at least at this
stage), mapping between the initial data sets for the field equations.
As we have seen above, this renormalization acts in such a way that at
each its step the deformed data satisfy the constraints. The time
evolution, in turn, of any such data set generates a one-parameter
family of solutions to the field equations which interpolates between
the initial inhomogeneous space-time to be averaged-out and its
renormalized (deformed) counterpart.

\subsection{The Ricci--Hamilton flow}

In order to replace the discrete operator ${\cal R}$, describing the
effect of renormalization of $g_{ab}$, with a continuous flow we can
start by discussing some geometrical implications of equation
(\ref{rg}). According to the Renormalization Group analysis of the
previous section, they follow by considering the average
$<f>_{\epsilon}$ as a functional of the metric and thinking of the
metric $g_{ab}$ as a running coupling constant, depending on the
cut-off. In this connection, it can be verified that we can
equivalently interpret (\ref{rg}) as obtained by considering the
variation of $<f>_{\epsilon}$ under a suitable smooth deformation of
the background metric $g_{ab}$, rather than by deforming the
(euclidean) radius of the balls $\{B(x_i,\epsilon)\}$.
\vskip 0.5 cm
As a matter of fact, we can equivalently rewrite the second term on
the right hand side of (\ref{rg}) as
\begin{equation}
\label{lin}
<R>_{\epsilon} <f>_{\epsilon} - <R f>_{\epsilon} = -
D<f>_{\epsilon}\cdot \frac{\partial g_{ab}}{\partial\eta},
\end{equation}
where $D<f>_{\epsilon}\cdot \partial g_{ab}/\partial \eta$ denotes the
formal linearization of the functional $<f>_{\epsilon}$ in the
direction of the symmetric $2-$tensor $\partial g_{ab}/ \partial
\eta$, and where
\begin{equation}
\label{ham}
\frac{\partial
g_{ab}(\eta)}{\partial\eta}=\frac{2}{3}<R(\eta)>
g_{ab}(\eta)-2R_{ab}(\eta),
\end{equation}
$R_{ab}(\eta)$ being the components of the Ricci tensor
$Ric(g(\eta))$, and $<R(\eta)>$ is the average scalar curvature given
by
\begin{equation}
<R(\eta)>=\frac{1}{vol(\Sigma)}\int_{\Sigma}R(\eta) d\mu_{\eta}.
\end{equation}
Indeed, the linearization of $<f>_{\epsilon}$ in the direction of the
generic $2-$tensor $\partial g_{ab}/\partial \eta$ is provided by

\begin{equation}
\label{gen}
 D<f>_{\epsilon}\cdot \frac{\partial g_{ab}}{\partial\eta}=
\frac{1}{2}<fg^{ab}\frac{\partial}
{\partial\eta}g_{ab}>_{\epsilon}-
\frac{1}{2}<f>_{\epsilon}<g^{ab}\frac{\partial}{\partial
\eta}
g_{ab}>_{\epsilon},
\end{equation}
so (\ref{lin}) follows, given the expression (\ref{ham}) for $\partial
g_{ab}/\partial \eta$.
\vskip 1 cm
According to what has been said in the previous paragraphs, the
effective distribution of matter sources, according to a set of
instantaneous observers $\{B(x_i,\epsilon)\}$, is characterized by the
underlying three-geometry thought of as an effective parameter
depending on the cutoff $\epsilon$.  Since the Hamiltonian and the
divergence constraints hold at each stage of the renormalization
procedure, (they fix the effective Hamiltonians which otherwise are
undetermined up to a constant factor), the renormalization of the
matter fields is intrinsically tied with the renormalization of the
three-metric.
\vskip 0.5 cm
The invariance of the long distance properties of the matter
distribution, under simultaneous change of the cutoff $\epsilon$ and
the parameter $g_{ab}$, can be expressed as a differential equation
for the effective Hamiltonian ${\cal H}(\rho,{\bf J})$, (actually for
the partition function associated with this effective Hamiltonian),
namely,
\be
\label{betaflow}
[-\epsilon\frac{\partial}{\partial\epsilon}+{\beta}_{ab}(g)
\frac{\partial}{\partial{g}_{ab}}]\sum_{\rho,J}
\exp[-{\cal H}(\rho,{\bf J})]=0.
\ee

Recall that ${\cal H}(\rho,{\bf J})$ is explicitly provided by (in the
given approximation)
\begin{eqnarray}
{\cal H}(\rho,{\bf J})= <\alpha p(\rho)+ {\alpha}^iJ_i>_{\epsilon}
\sum_h vol(B(x_h,\epsilon)),
\end{eqnarray}
where the average $<\alpha p(\rho)+ {\alpha}^iJ_i>_{\epsilon}$ is a
functional of the three-metric $g_{ab}$, here thought of as the
running coupling constant.

Thus, in order that the equation (\ref{betaflow}) is satisfied, it is
{\it sufficient} that
\be
\label{firecrackers}
[-\epsilon\frac{\partial}{\partial\epsilon}+{\beta}_{ab}(g)
\frac{\partial}{\partial{g}_{ab}}]
<\alpha p(\rho)+ {\alpha}^iJ_i>_{\epsilon}=0.
\ee

The Renormalization Group equation (\ref{firecrackers}) states that
increasing the cutoff length (i.e., the radius of the averaging balls)
from $\epsilon$ to $e^{\gamma}\epsilon$, while deforming the metric
$g_{ab}$ by flowing along the beta function ${\beta}_{ab}(g)$ for a
parameter-time $\gamma$, has no net effect on the long distance
properties of the considered system.
\vskip 0.5 cm
According to the above remarks, and equation (\ref{empirical}), it can
be verified that the beta function yielding for (\ref{firecrackers}),
defined by
\begin{eqnarray}
\epsilon\frac{\partial}{\partial\epsilon} g_{ab}=
{\beta}_{ab}(g),
\end{eqnarray}
is exactly provided by (\ref{ham}), namely,

\begin{equation}
{\beta}_{ab}(g)=\frac{\partial
g_{ab}(\eta)}{\partial\eta}=\frac{2}{3}<R(\eta)>
g_{ab}(\eta)-2R_{ab}(\eta),
\end{equation}
where the parameter $\eta$ is the logarithmic change of the cutoff
length $\epsilon$.
\vskip 0.5 cm
Since the manifold $\Sigma$ is compact, (\ref{ham}) has to be
interpreted as a Renormalization Group equation in a finite geometry,
and thus the relevant phenomena are here related to finite size
scaling (see e.g. \cite{Cardy}).  A continuous theory, describing the
(universal) properties of the cosmological sources and of the
corresponding geometry, may arise when the correlation length
associated with the distribution of cosmological matter is of the
order of the size of the underlying manifold.

\vskip 1 cm

The metric flow (\ref{ham}) is known as the Ricci-Hamilton flow
\cite{Ham}, studied in connection with the quasi-parabolic flows on
manifolds; quite independently it has been discussed in investigating
the Renormalization Group flow for general $\sigma$-models (see e.g.
\cite{Lott} and references quoted therein).  The Ricci-Hamilton flow
is always solvable \cite{Ham} for sufficiently small $\eta$ and has a
number of useful properties, (apart from being volume preserving which
is simply a consequence of the normalization chosen), namely, any
symmetries of $g_{ab}(\eta_o)$ are preserved along the $g_{ab}(\eta)$
flow for all $\eta> \eta_o$, and the limiting metric (if attained)
${\overline{g}}_{ab}=\lim_{\eta \rightarrow \infty} g_{ab}(\eta)$ has
constant sectional curvature.  Thus equation (\ref{ham}), with the
initial condition $g_{ab}(0)=g_{ab}$, defines (when globally solvable)
a smooth family of deformations of the initial three-manifold,
deforming it into a three-space of constant curvature.

\subsection{Fixed points and basins of attraction}

Thus the point of the above discussion is that in order to arrive at a
{\it fixed point} of the RG equation (\ref{rg}), the geometry has to
be deformed according to the Ricci-Hamilton flow (\ref{ham}).  In this
setting of the problem, the Ricci-Hamilton equations appear naturally
and in fact the approach proposed, enables us to attach a physical
meaning to them within the coarse-graining picture. This element was
lacking in \cite{Mauro} where Hamilton's theorem appeared rather {\it
ad hoc}.  On the other hand, our approach demonstrates that the
smoothing issue is deeply connected with the geometry and exhibits how
this relationship works.
\bigskip
First some general remarks.\par A fixed point is a point in the
coupling constants space that satisfies
\be
g_{ab}^{\star}= {\cal R} (g_{ab}^{\star}),
\label{fixed}
\ee
i.e. it is mapped onto itself by RG transformation.

Under RG transformation length scales are reduced by a factor $m+1$.
Namely, for the block variables, the correlation length measured in
units appropriate for them, $L_m(\epsilon_m)$, is smaller than the
correlation length $L_o$ of the initial system measured in units of
$\epsilon_o$.  The actual physical value of the correlation length $L$
is of course unchanged by the process of blocking, thus
$L=L_m(\epsilon_m)=L_o
\epsilon_o$,
so
\be
L_m=\frac{L_o}{m+1}.
\ee
Since $L_m<L_o$, the system with Hamiltonian ${\cal H}^{(m)}$ must be
further >from criticality than the original system. Thus, we conclude
that the system is at a new effective reduced ``temperature"
$g_{ab}^{(m)}$. At a fixed point (\ref{fixed}), $L^\star$ can be only
zero or infinity since then $L^\star=L^\star/(m+1)$.

As is standard in RG analysis, we will refer to a fixed point with
$L=\infty$ as a {\it critical fixed point}, when $L=0$ we will call it
{\it trivial}. Each fixed point has its domain or {\em basin of
attraction}, namely, the points in the coupling constants space in
such a basin necessarily flow towards and end up at the fixed point,
after an infinite number of iterations of RG transformation.
\bigskip

Let us, for the purpose of clarity, employ for a moment the
ferromagnetic analogy. In this case, for a system exhibiting a phase
transition, there are two attractive fixed points. One is the
high-temperature fixed point which attracts each point with
$T>T_{crit}$ in the coupling constants space, and it corresponds to
the effective Hamiltonian for the system as $T\rightarrow \infty$. In
this phase the variables assume random values and are uncorrelated.
Upon a sensible blocking of such a system the probability distribution
of the block variables remains unchanged.

The second fixed point is the low-temperature fixed point which is the
effective Hamiltonian for the system when $T\rightarrow 0$. This
corresponds to a system in a complete spins alinement and the block
variables are ordered then. Every point corresponding to $T<T_{crit}$
is eventually attracted to this fixed point.
\bigskip

Across the Hamiltonian space there should exist then a surface, the
so-called {\it critical surface}, which separates the effective
Hamiltonians flowing to the high-temperature fixed point from those
flowing to the low-temperature fixed point. Notice that the word {\it
surface} here has rather a heuristic meaning. Indeed, the {\it set} of
metrics, separating those flowing towards a ``low-temperature'' fixed
points >from those flowing towards a ``high-temperature'' fixed point,
has quite a complex structure whose understanding is deeply connected
with some, yet unsolved, conjectures in $3D$-manifold topology, (see
below for more details).
\vskip 0.5 cm
If we now choose to start with a point on the critical surface, then
upon RG transformation it will stay within the critical surface.
There is a possibility that, as the number of iterations of RG goes to
infinity, the Hamiltonian will tend to a finite limit ${\cal
H}^\star$.  This point is the critical fixed point and within the
critical surface it is attractive, (this is roughly speaking the basic
mechanism for universality), along the direction out of the critical
surface it is repulsive. This fixed point is related to the singular
critical behaviour of the system due to the fact that all points in
its basin of attraction have infinite correlation length. The simplest
case is when the fixed points are isolated points, but it is also
possible to have lines or surfaces of fixed points.
\vskip 0.5 cm
With these preliminary remarks along the way, let us discuss how some
of the above general characteristic of the Renormalization Group flows
find their proper counterpart in the particular situation we are
analysing.
\vskip 0.5 cm
In the previous paragraph we showed that it was possible to replace
the discrete operation of increasing the scale size of the
observational averaging region by a ``transformation", smoothly
deforming the background metric $g_{ab}$, which turned out to be the
Ricci-Hamilton flow. In this setting, following the example above, we
would like to adopt the fundamental hypothesis linking RG to the
critical phenomena, namely, the existence of a ``critical" metric (on
the critical surface) $g_{ab}^{crit}$, and of a ``fixed point" metric
$g_{ab}^\star$, such that
\be
\lim_{m\rightarrow \infty}{\cal R}^{(m)}(g_{ab}^{crit})=
g_{ab}^{\star}.
\label{crit}
\ee
\vskip 0.5 cm
In (\ref{crit}) $g_{ab}^\star$ is a mathematical object invariant
under RG and we assume that $g_{ab}^{crit}$ represents the physics (in
a sense to be clarified further) of curvature fluctuations of a
manifold at its critical point (``critical manifold'').  Since the
Ricci-Hamilton flow can be interpreted as a dynamical system on a set
of closed riemannian manifolds, we can adopt the following
interpretation of (\ref{crit}).  Suppose, that we look at our manifold
through a ``microscope" and are able to discern the curvature
fluctuations down to a size $\epsilon_m$.  Then ${\cal R}^{(m)}$
represents the operation of decreasing the magnification factor, by
$m$ say, i.e. the sample seen appears to shrink by this factor. We
have to assume that the system is sufficiently large so that the edges
of the sample will not appear in the view.  The hypothesis
(\ref{crit}) states that if we decrease the magnification by a
sufficiently large amount, we shall not see any change if we decrease
it even further.
\vskip 0.5 cm

We are going to describe to what extent such hypothesis holds.
\vskip 0.5 cm

We already said that the Ricci-Hamilton flow (\ref{ham}), while always
solvable for sufficiently small $\eta$ \cite{Ham}, may not yield for a
non-singular solution as $\eta$ increases. Hamilton noticed that there
are patterns in the kind of singularities that may develop.
Typically, the curvature blows up but in a very regular way (e.g. for
$S^1\times S^2$ with the standard symmetric metric).  This has led him
to a research program which, roughly speaking, amounts to saying that
any three-manifold can be decomposed into pieces on which the
Ricci-Hamilton flow is global and thereby, each of these pieces can be
smoothly deformed into a locally homogeneous three-manifold.
Singularities may develop in the regions connecting the smooth-able
pieces, but such singularities should be of a finite number of types
and all of a rather symmetric nature (namely, if they are blown up,
they should be associated with symmetric manifolds as e.g. $S^1\times
S^2$
\cite{CIJ, Jackson}).
\bigskip

It may be said that Hamilton's program is an analytic approach to
prove Thurston's conjecture, which claims that any closed
three-manifold can be cut into pieces, such that each of them admits a
locally homogeneous geometry. The rationale, underlying this {\it
analytical} program towards Thurston's geometrization conjecture, lies
in the above nice structural properties of the Ricci-Hamilton flow.
Several steps are involved in this program. Let us briefly recall them
since, even if some of them are yet unproven, they shed light on the
assumption (\ref{crit}) and on (\ref{ham}) when interpreted as a
Renormalization Group flow.
\vskip 0.5 cm

The {\bf first step} is the assignment of an {\it arbitrary} metric
$g$ on the three-manifold $\Sigma$.  In the Renormalization Group
approach, this corresponds to picking up a vastly inhomogeneous and
anisotropic geometry for the physical space, (here equivalent to the
{\it high-temperature phase}).  Such a choice may not conform to the
actual {\it quasi-homogeneous} three-geometry of the physical space as
is experienced now. However this quasi-homogeneity, in our opinion,
may be related to the possibility that the actual universe is near
criticality, a circumstance that we want to discuss rather than assume
from the outset.
\vskip 0.5 cm
The {\bf second step} is to deform this metric $g$ via the
Ricci-Hamilton flow (\ref{ham}).  In general, this flow develops local
singularities which should be related to the manifold decomposition in
Thurston's conjecture. Away from each of the local singularities it is
conjectured (but not yet proved) that the Ricci-Hamilton flow
approaches that of a locally homogeneous geometry in each disconnected
piece.\par

This picture may be consistent with our blocking procedure yielding
for (\ref{ham}) as the Renormalization Group flow.  According to the
analysis, carried out in section 3.1, a highly inhomogeneous and
anisotropic geometry $(\Sigma, g)$ can be characterized by minimal
geodesic ball coverings $\{B(x_i,\epsilon_o)\}$, whose balls are to a
good approximation largely uncorrelated when seen on a suitable scale.
This means that the values of the scalar curvatures, $R(i)$, evaluated
at the centers of the balls of the covering, are random variables.
Upon enlarging the balls and rescaling, certain regions of the
manifold might be such that correlations develop among the
corresponding $R(i)$, whereas in other regions, no matter how we
enlarge the balls and rescale, the $R(i)$ will remain independently
distributed random variables. The former regions, exhibiting a
persistence length, are those that should approach a locally
homogeneous geometry under blocking and rescaling (i.e., under the
flow (\ref{ham})). The latter regions should rather develop
singularities under (\ref{ham}), since no matter how much we block and
rescale, the curvature will maintain its {\it white noise} character.

\vskip 0.5 cm
The {\bf third step} is to study the behavior of (\ref{ham}) for the
locally homogeneous geometries, for this accounts for the structure of
the critical set of (\ref{ham}). This has been accomplished
\cite{Jackson}, and one finds that, depending on the initial locally
homogeneous geometry, the Ricci-Hamilton flows either, {\it (i)}
converges to a constant curvature metric, {\it (ii)} asymptotically
approaches (as $\eta\to\infty$) a flat degenerate geometry, of either
two or one dimensions ({\it pancake} or {\it cigar} degeneracy), with
the curvature decaying at the rate $1/\eta$, or, {\it (iii)} hits a
curvature singularity in finite time, with this singularity being that
of the Ricci-Hamilton flow for the standard metric on $S^2\times S^1$.
Note that constant curvature geometries always occur whenever the
manifold can support them, (in dimension three, constant curvature
manifolds and Einstein manifolds are synonymous). It is also quite
interesting to note that the Ricci-Hamilton flow of homogeneous
metrics usually (with a few exceptions) tends to approach or converge
to the maximally symmetric homogeneous metric in the class considered
(see \cite{Jackson} for details).
\vskip 0.5 cm
Interpreting (\ref{ham}) as the Renormalization Group flow, it follows
that locally homogeneous geometries evolving under (\ref{ham}) towards
an isolated constant curvature manifold, are sinks describing a stable
phase of the corresponding cosmological model. For instance a FLRW
model (with closed spatial sections) characterizes such a phase.
Non-isolated constant curvature manifolds (e.g. flat three-tori)
provide less trivial examples of limiting behaviour of (\ref{ham}),
(see e.g. \cite{RH, CIJ}). The locally homogeneous manifolds
non-admitting Einstein manifolds (i.e.  there are no left-invariant
Einstein metrics on the group $SL(2,{\rm I\!R})$), provide even more
interesting behavior.  In this case, a metric renormalized under the
action of (\ref{ham}) develops degeneracies and one gets, in our
setting, an {\it effective} cosmological model with spatial sections
of lower dimensionality.\par
\vskip 0.5 cm
All these limit points of (\ref{ham}), either fixed or not, have their
own basins of attraction. Rigorously speaking, they are the sets of
three-metrics flowing under (\ref{ham}) to the respective limit
points, briefly discussed above. There are nine such basins of
attraction, corresponding to the nine classes of homogeneous
geometries that can be used to model ({\it by passing to the universal
cover}) the local inhomogeneous geometries on closed three-manifolds.
By labelling these classes according to the minimal isometry group of
the geometries considered we distinguish the following basins (here we
follow closely the exposition in \cite{Jackson}):\par
\vskip 0.5 cm
{\it (i): The ${\rm I\!R}^3$-Basin}. It contains all three-metrics
flowing towards the homogeneous flat ${\rm I\!R}^3$ metrics. This basin
is eventually attracted by flat space, (flat tori, when reverting to
the original manifold rather than to its universal cover).\par
\vskip 0.5 cm
{\it (ii): The $SU(2)$-Basin}. It contains all three-metrics flowing
towards the three-parameter family of homogeneous $SU(2)$ metrics.
This class admits Einstein metrics, in particular the round metrics on
the three-sphere. This basin is exponentially attracted to the round
three-sphere, (modulo identifications). It is the basin of attraction
yielding for closed FLRW cosmological models. \par
\vskip 0.5 cm
{\it (iii): The $SL(2,{\rm I\!R})$-Basin}. It contains all three-metrics
flowing towards the three-parameter family of homogeneous $SL(2,
{\rm I\!R})$ metrics.  This class does not admit Einstein metrics. This basin
goes degenerate, yielding for a {\it pancake} degeneracy whereby a
two-dimensional geometry survives: two of the components of the metric
increase without bound while the other shrinks to zero.
\par
\vskip 0.5 cm
{\it (iv): The Heisenberg-Basin}. It contains all three-metrics
flowing towards the three-parameter family of homogeneous Nil-metrics.
Again, this class does not contain any Einstein metrics. This basin
too undergoes a pancake degeneracy.\par
\vskip 0.5 cm
{\it (v): The $E(1,1)$-Basin}, where $E(1,1)$ is the group of
isometries of the plane with flat Lorentz metric.  It contains all
three-metrics flowing towards the three-parameter family of
homogeneous Solv-metrics. Also this basin fails to contain Einstein
metrics. This basin eventually exhibits a {\it cigar} degeneracy: the
curvature dies away, and while one diameter expands without bound, the
other two diameters shrink to zero.\par
\vskip  0.5 cm
{\it (vi): The $E(2)$-Basin}, where $E(2)$ is the group of isometries
of the euclidean plane. It contains all three-metrics flowing towards
the three-parameters family of homogeneous Solv-metrics containing the
flat geometry. This basin is eventually attracted by flat metrics.\par
\vskip 0.5 cm
{\it (vii): The $H(3)$-Basin}, where $H(3)$ is the group of isometries
of hyperbolic three-space. It contains all three-metrics flowing
towards the one-parameter family of homogeneous metrics constant
multiples of the standard hyperbolic metric.  This basin is attracted
to hyperbolic space.\par
\vskip 0.5 cm
{\it (viii): The $SO(3)\times {\rm I\!R}^1$-Basin}. It contains all
three-metrics flowing towards the two-parameter family of homogeneous
metrics obtained by rescaling the standard product metric on
$S^2\times {\rm I\!R}^1$.  It does not contain Einstein metrics. This is
a singular basin, it is attracted towards a curvature singularity: the
round two-sphere shrinks while the scale on ${\rm I\!R}^1$, (or if you
prefer, the $S^1$ factor in the original manifold), expands.\par
\vskip 0.5 cm
{\it (ix): The $H^2\times {\rm I\!R}^1$-Basin}, where $H(2)$ is the
group of isometries of the hyperbolic plane. It contains all
three-metrics flowing towards the two-parameter family of homogeneous
metrics obtained by rescaling the product metric on the product
manifold, ${\rm I\!R}^1\times H^2$. Again, this basin does not contain
Einstein manifolds, and it is attracted towards a pancake degeneracy.
\vskip 1 cm
The basins of attraction just described, and in particular those
yielding for fixed points (Einstein manifolds), are relatively
uninteresting in connection with the Renormalization Group
interpretation of (\ref{ham}). Such fixed points, e.g.  the round
three-sphere, or the flat three-tori, are all totally attractive. As
already recalled they can be thought of as distinct stable phases
yielding for distinct cosmological models, characterized by the nature
of the metric (and of its infinitesimal deformations) at the fixed
point. For instance, the $SU(2)$-Basin characterizes FLRW models with
closed spatial sections, and the related homogeneous anisotropic
models, (see section 6 for more details).
\vskip 0.5 cm

\subsection{Critical fixed points: an example}
\label{examplecrit}

The existence of critical fixed points, (critical in the sense of
(\ref{crit})), characterizing the (universal) properties of
(continuous) phase transitions between two different cosmological
regimes, cannot be immediately read off from the above analysis. It is
rather the consequence of the (conjectured) existence of the
decomposition of a manifold into pieces that is associated, according
to Hamilton, to the local singularities of the flow (\ref{ham}), (see
the {\it second step} of Hamilton-Thurston geometrization program).
The origin of this connection between critical behavior of
(\ref{ham}), in the sense of Renormalization Group, and Hamilton's
program can be seen by considering the following detailed example.
\vskip 0.5 cm

Let us assume that topologically $\Sigma$ is a three-sphere,
$\Sigma\simeq S^3$.  We shall consider on $\Sigma$ a metric $g_1$
obtained by glueing through a smooth connected sum two copies of a
round three-sphere
\begin{eqnarray}
\Sigma\simeq S^3_{(1)}\sharp S^3_{(2)},
\end{eqnarray}
and endowing each $S^3_{(i)}$, $i=1,2$, factor with a round metric of
volume $v_1= 1$, and the joining tube
\begin{eqnarray}
S^2\times ([0,1]\subset {\rm I\!R}^1),
\end{eqnarray}
with the standard product metric of volume $const. e^{-3\tau}$,
($\tau$ is a suitable parameter, see below), for $\tau \gg 1$.\par
\vskip 0.5 cm
To explicitly construct this latter metric we can proceed as
follows:\par
\vskip 0.5 cm
Let $y_1$ and $y_2$ respectively denote two chosen points in both
factor copies, $S^3_{(1)}$ and $S^3_{(2)}$.  Let $h_{(i)}\colon
{\rm I\!R}^3\to S^3_{(i)}$, $i=1,2$, be two imbeddings given by the
exponential mappings
\begin{eqnarray}
\exp_{y_1} &\colon & T_{y_1}S^3_{(1)}\simeq {\rm I\!R}^3\to
S^3_{(1)},\nonumber\\
\exp_{y_2} &\colon & T_{y_2}S^3_{(2)}\simeq {\rm I\!R}^3\to
S^3_{(2)}.
\end{eqnarray}

We assume that $h_{(1)}$ preserves the orientation while $h_{(2)}$
reverses it.\par Let $\alpha\colon (0,\infty)\to (0,\infty)$ denote an
orientation reversing diffeomorphism, and define $\alpha_3\colon {\rm I\!
R}^3/\{0\}\to {\rm I\!R}^3/\{0\}$ by the map
$\alpha_3(v)=\alpha(|v|)\frac{v}{|v|}$ for every vector $v\in {\rm I\!
R}^3/\{0\}$. For every point $x_1=h_1(v)$ in the geodesic ball
$B(y_1,r)/\{0\}\subset S^3_{(1)}$, with $0< r\leq\frac{\pi}{2}$, (with
$\pi/2$ the injectivity radius of the unit three-sphere $S^3_{(i)}$),
we identify $h_1(v)$ with $h_2(\alpha_3(v))$ $\in$
$B(y_2,r)/\{0\}\subset S^3_{(2)}$.\par
\vskip 0.5 cm
The space obtained in this way
\begin{eqnarray}
S^3_{(1)}\sharp S^3_{(2)}= (S^3_{(1)}-\{y_1\})
\cup_{h_{(2)}\alpha_3 h_{(1)}^{-1}}(S^3_{(2)}-\{y_2\})
\end{eqnarray}

is a particular realization of the connected sum \cite{Kosi} of two
copies of unit three-spheres.\par
\vskip 0.5 cm
In order to give to the neck, joining the two $S^3_{(i)}$, a
cylindrical shape we blow up \cite{Uhle} the metrics of the
three-spheres in the neighborhoods of the points $y_1$ and $y_2$.
Consider, for simplicity, only the $S^3_{(1)}$ factor since the
argument goes in an analogous way also for the $S^3_{(2)}$ factor.
\vskip 0.5 cm
Exploiting the exponential mapping the round metric of $S^3_{(1)}$,
(actually {\it any} sufficiently smooth metric), can be written in a
neighborhood of $y_1$ in {\it geodesic polar coordinates} as
\begin{eqnarray}
g(x)= dr^2 + r^2h_{ij}d\theta^i d\theta^j +{\cal O}(r^4),
\end{eqnarray}
where $r={\rm dist}(x,y_1)$ is the distance between $y_1$ and the
point considered, $h_{ij}d\theta^i d\theta^j$ is the metric on the
two-dimensional unit sphere $S^2$, and as usual, the higher order
correction terms involve the curvature. \par
\vskip 0.5 cm
Now, if we blow up this metric by rescaling it through $r^2$ we get,
up to curvature corrections, a distorted cylindrical metric
\begin{eqnarray}
{\tilde g}(x)=\frac{g(x)}{r^2}=
\frac{dr^2}{r^2} + h_{ij}d\theta^i d\theta^j +{\cal O}(r^2).
\end{eqnarray}
\vskip 0.5 cm

In order to eliminate the axial distortion due to $\frac{dr^2}{r^2}$,
we substitute for the radial variable $r$, introducing a new
coordinate $\tau$ defined as
\begin{eqnarray}
r=\exp [-\tau].
\end{eqnarray}

When expressed in terms of $\tau$ we get that the blown up metric
${\tilde g}(x)$ reduces to
\begin{eqnarray}
{\tilde g}(x)= d\tau^2 + h_{ij}d\theta^i d\theta^j +{\cal
O}(e^{-2\tau}).
\end{eqnarray}
\vskip 0.5 cm
Thus, the blown up metric approaches the cylindrical metric
exponentially fast as $\tau\to\infty$. In order to introduce smoothly
such a metric on the neck of $S^3_{(1)}\sharp S^3_{(2)}$ we can
proceed as follows.
\vskip 0.5 cm
Choose smooth functions $\delta_{(i)}(r)$, $i=1,2$, satisfying
\begin{eqnarray}
\delta_{(i)}(r_i) &= & \frac{1}{r_i^2}\,,\, \; 0<r_i\leq
\frac{\pi}{4}\,,\nonumber\\
\delta_{(i)}(r_i) &= & C^{\infty}\,,\,\; {\rm decreasing}, \:
\frac{\pi}{4}\leq r_i\leq \frac{\pi}{2}\,,\\
\delta_{(i)}(r_i) &= & 1\,,\, \; r_i\geq \frac{\pi}{2}\,,\nonumber
\end{eqnarray}
where $r_i= {\rm dist}(x,y_{(i)})$. Also, let us introduce the
following inclusion maps
\begin{eqnarray}
{\chi}_{(1)}&\colon & S^3_{(1)}-\{y_1\}\hookrightarrow
(S^3_{(1)}-\{y_1\})
\cup_{h_{(2)}\alpha_3 h_{(1)}^{-1}}(S^3_{(2)}-\{y_2\}),
\nonumber\\
{\chi}_{(2)}&\colon & S^3_{(2)}-\{y_2\}\hookrightarrow
(S^3_{(1)}-\{y_1\})
\cup_{h_{(2)}\alpha_3 h_{(1)}^{-1}}(S^3_{(2)}-\{y_2\}),
\\
{\chi}_{(3)}&\colon & h_1(T_{y_1}S^3_{(1)}-\{0\})\cup
h_2(T_{y_2}S^3_{(2)}-\{0\})
\hookrightarrow   (S^3_{(1)}-\{y_1\})
\cup_{h_{(2)}\alpha_3 h_{(1)}^{-1}}(S^3_{(2)}-\{y_2\}),\nonumber
\end{eqnarray}
($\chi_{(1)}$ and $\chi_{(2)}$ denote the inclusions of the spheres,
minus the points $y_{(i)}$, into the connected sum; $\chi_{(3)}$ is
the inclusion of the neck).
\vskip 0.5 cm
If $\{\xi_{\alpha}\}$ is a partition of unity associated with the
covering corresponding to the above inclusions $\chi_{(\alpha)}$, with
$\alpha=1,2,3$, we can then define the following metric on
$S^3_{(1)}\sharp S^3_{(2)}$,

\be
\label{symmetric}
{\tilde g}(x) =\sum_{\alpha}\xi_{\alpha} {\chi}^*_{(\alpha)}[g(x)\cdot
\delta_{(i)}({\rm dist}(x,y_{(i)}))].
\ee
\vskip 0.5 cm

For $x$ not in the geodesic balls (of radius $\pi/2$), centered on
$y_1$ and $y_2$, this is the standard round metric on the unit
three-sphere; for $x$ in the geodesic balls of radius $\pi/4$,
centered on $y_1$ and $y_2$, this is, up to curvature corrections, the
cylindrical metric introduced above. For $x$ in the annuli between
such balls ${\tilde g}$ is a smooth interpolating metric joining the
spheres to the cylindrical neck.\par
\vskip 0.5 cm
By expressing (\ref{symmetric}) in terms of the variable $\tau$, we
get the metric on $S^3_{(1)}\sharp S^3_{(2)}$ which is the round
metric on each $S^3_{(i)}/B(y_i,e^{-\tau})$ factor, and these factors
are connected, for $\tau$ large enough (i.e., nearby $y_i$), by a thin
flat cylinder.

\vskip 1 cm

The Ricci-Hamilton evolution of $(S^3_{(1)}\sharp S^3_{(2)}, {\tilde
g})$ can be explicitly constructed as follows.
\vskip 0.5 cm
According to \cite{Jackson} let us write the metric on $S^2\times
{\rm I\!R}^1$ as

\begin{eqnarray}
{\tilde g}|_{{\rm neck\, of}\,S^3_{(1)}\sharp S^3_{(2)}}= Dg_{{\rm I\!
R}^1}+Eg_{S^2},
\end{eqnarray}
where $g_{{\rm I\!R}^1}$ is the metric on ${\rm I\!R}^1$, $g_{S^2}$ is the
round metric on the unit two-sphere, and $D$ and $E$ are constants.
The Ricci-Hamilton flow (\ref{ham}) preserves the structure of this
metric and reduces to the coupled system of ordinary differential
equations
\begin{eqnarray}
\frac{d}{d\eta}E &= & -\frac{2}{3},\nonumber\\
\frac{d}{d\eta}D &= & \frac{4}{3}\left( \frac{D}{E} \right),
\end{eqnarray}

\noindent which immediately integrate to

\begin{eqnarray}
E &= & E_0-\frac{2}{3}\eta,\nonumber\\ D &= &
\frac{D_0E_0^2}{[E_0-(2/3)\eta]^2},
\label{jimandjack}
\end{eqnarray}

\noindent where $E_0^2=E^2(\eta=0)$ is the initial radius of the
round two-sphere, whereas $D_0=D(\eta=0)$ is the scale on ${\rm I\!
R}^1$.
\vskip 0.5 cm

Given this solution of (\ref{ham}), we can construct the
Ricci-Hamilton evolution of $(S^3_{(1)}\sharp S^3_{(2)}, {\tilde g})$
by looking for a solution of (\ref{ham}) in the form

\be
\label{singmet}
{\tilde g}(x;\eta) =\sum_{\alpha}\xi_{\alpha}
{\chi}^*_{(\alpha)}(\eta)[g(x)\cdot
\delta_{(i)}({\rm dist}_{\eta}(x,y_{(i)}))],
\ee
where the inclusion maps ${\chi}_{(\alpha)}$ depend now on the
deformation parameter $\eta$. We assume that the metric $g(x)$, (the
round metric on the spheres $S^3_{(i)}$ and the standard product
metric on the neck$\simeq S^2\times{\rm I\!R}^1$), being locally
homogeneous, is preserved by (\ref{ham}) since the Ricci-Hamilton flow
preserves isometries.  As the {\it plumbing} between the spheres and
the neck shrinks as $\eta$ increases, (as the $S^2$ factor in the
neck), the inclusions ${\chi}_{(\alpha)}$ are necessarily
$\eta$-dependent.  The dynamics of ${\chi}_{(\alpha)}$ can be obtained
as follows.
\vskip 0.5 cm
The Ricci-Hamilton flow (\ref{ham}) for ${\tilde g}(x;\eta)$ is given
by
\begin{eqnarray}
\left\{ \begin{array}{ll}
\frac{\partial
{\tilde g}_{ab}(\eta)}{\partial\eta}&= \frac{2}{3} <{\tilde R}(\eta)>
{\tilde g}_{ab}(\eta)-2{\tilde R}_{ab}(\eta) \nonumber\\ {\tilde
g}_{ab}(\eta=0)&= {\tilde g}_{ab},
\end{array}\right.
\label{ham2}
\end{eqnarray}
and, in terms of the pulled-back metric it is
\begin{eqnarray}
\frac{\partial}{\partial\eta}
[{\chi}^*_{(\alpha)}(\eta)g(x)\cdot
\delta_{(i)}
]_{ab}&= & \frac{2}{3} <{\tilde R}(\eta)>
[{\chi}^*_{(\alpha)}(\eta)g(x)\cdot
\delta_{(i)}
]_{ab}\nonumber\\ &- &2{\tilde R}({\chi}^*_{(\alpha)}(\eta)g(x)\cdot
\delta_{(i)}
)_{ab}.
\label{ham3}
\end{eqnarray}
\vskip 0.5 cm
\noindent The left hand side of (\ref{ham3}) can be evaluated
according to a suggestion first exploited by D. DeTurck, {\it viz.},
\vskip 0.5 cm
\begin{eqnarray}
\frac{\partial}{\partial\eta}
[{\chi}^*_{(\alpha)}(\eta)g\cdot
\delta_{(i)}
]_{ab}(x)&= & {\chi}^*_{(\alpha)}(\eta)\left[
\frac{\partial}{\partial\eta}
[g\cdot
\delta_{(i)}
]_{ab}({\chi}(\eta,x)) \right]\nonumber\\ &+
&{\chi}^*_{(\alpha)}(\eta)
\left[ L_{w(\eta)}[g\cdot
\delta_{(i)}
]_{ab}({\chi}(\eta,x)) \right],
\end{eqnarray}
\vskip 0.5 cm
\noindent where the Lie derivative  $L_{w(\eta)}[g\cdot
\delta_{(i)}]_{ab}({\chi}(\eta,x))$ is along the vector
field $w(\eta;\alpha)$ which generates the $\eta$-evolution of the
inclusions ${\chi}_{(\alpha)}$, {\it viz.},
\begin{eqnarray}
\frac{\partial
{\chi}_{(\alpha)}(\eta)}{\partial\eta} &= & w(\eta;
{\chi}_{(\alpha)}(\eta)),\nonumber\\ {\chi}_{(\alpha)}(\eta=0) &= &
{\chi}_{(\alpha)}.
\end{eqnarray}
\vskip 0.5 cm

Let us denote by $<R>_{(\alpha)}$ the average of the scalar curvature
$R(\chi_{(\alpha)}(x))$ over the images of the inclusions
$\chi_{(\alpha)}$, ({\it viz.}, for $\alpha=1,2$, $<R>_{(\alpha)}$ is
the average over $S^3_{(i)}$, while for $\alpha=3$, the average is
over the neck). In terms of these averages, the Ricci-Hamilton flow
(\ref{ham3}) can be written as

\begin{eqnarray}
\frac{\partial}{\partial\eta}
[\delta_{(i)}\cdot g_{ab}(\chi_{(\alpha)}(x))]&= &\frac{2}{3} < R
(\eta)>_{(\alpha)} {\delta_{(i)}\cdot g}_{ab}(\chi_{(\alpha)}(x)) -2
R_{ab}(\chi_{(\alpha)}(x))\nonumber\\ &+ & \frac{2}{3}
{\delta_{(i)}\cdot g}_{ab}(\chi_{(\alpha)}(x))
\left[<R(\eta)>-<R(\eta)>_{(\alpha)} \right]\nonumber\\
&- & L_{w(\eta)}[g\cdot
\delta_{(i)}]_{ab}({\chi}(\eta,x)).
\label{ham4}
\end{eqnarray}
\vskip 0.5 cm
As can be checked, the Ricci-Hamilton flow preserves the local
homogeneous geometries over the spheres $S^3_{(i)}$, and on the neck,
if and only if the vector field $w(\eta;\alpha)$ satisfies
\be
\label{vectorpart}
\frac{2}{3}
{\delta_{(i)}\cdot g}_{ab}(\chi_{(\alpha)}(x))
\left[<R(\eta)>-<R(\eta)>_{(\alpha)} \right]=
L_{w(\eta)}[g\cdot
\delta_{(i)}]_{ab}({\chi}(\eta,x)).
\ee
\vskip 0.5 cm
By taking the trace of (\ref{vectorpart}) with respect to
$g\cdot\delta_{(i)}$, we get

\begin{eqnarray}
<R>-<R>_{(\alpha)}={\bar{\nabla}}^k w_k(\eta;\alpha),
\end{eqnarray}
\vskip 0.5 cm
\noindent where ${\bar{\nabla}}$ denotes the riemannian connection
with respect to $g\cdot\delta_{(i)}$.
\vskip 0.5 cm
For $\alpha=1,2$, i.e. for the punctured spheres $S^3_{(i)}-\{y_i\}$,
the above relation yields upon integration over $S^3_{(i)}-B(y_i,r)$

\be
\label{blow}
\left[ <R>-<R>_{(\alpha)} \right]
vol \left( S^3_{(i)}-B(y_i,r) \right)=
\int_{S^2(\eta)}w_k(\eta;\alpha) d{\sigma}^k,
\ee

\noindent where $S^2(\eta)$ is the $\eta$-dependent boundary of
$S^3_{(i)}-B(y_i,r)$. Since $<R>-<R>_{(\alpha)}$, for $\alpha=1,2$, is
proportional to the average scalar curvature of the neck, the term on
the left hand side of (\ref{blow}) blows up as $(E_0-
\frac{2}{3}\eta)^{-1}$ as $\eta$ increases. On the other hand, by the
$S^2(\eta)$-
rotational symmetry, the term $w_k(\eta;\alpha)d{\sigma}^k$ is
spatially constant on the two-sphere boundary $S^2(\eta)$ of the
punctured three-spheres. Thus, (\ref{blow}) implies that as $\eta$
increases, the surface area of $S^2(\eta)$ shrinks to zero, as
$(E_0-\frac{2}{3}\eta)^{-1}$, by moving along the outer normal
direction of $S^2(\eta)\subset (S^3_{(i)}- B(y_i,r(\eta)))$.
\vskip 0.5 cm
For $\alpha=3$, i.e., for the neck, (\ref{blow}) yields
\be
\label{blow2}
\left[ <R>-<R>_{(\alpha=3)} \right]
vol \left( S^2(\eta)\times {\rm I\!R}^1 \right)=
\int_{S_{(2)}^2(\eta)}w_k(\eta;\alpha) d{\sigma}^k-
\int_{S_{(1)}^2(\eta)}w_k(\eta;\alpha) d{\sigma}^k,
\ee

\noindent where $S^2_{(2)}(\eta)-S^2_{(1)}(\eta)$
is the oriented boundary of the cylinder $S^2(\eta)\times {\rm I\!R}^1$.
Since $<R>-<R>_{(\alpha=3)}$ is, up to small correction terms (coming
from the collars joining the spheres to the neck), the average
curvature over the spheres $S^3_{(i)}$, $i=1,2$, (\ref{blow2}) simply
tells us that as the $S^2(\eta)$ factor in the neck shrinks, the
scale-length of the ${\rm I\!R}^1$ factor grows, so that the volume $vol
\left( S^2(\eta)\times {\rm I\!R}^1 \right)$ remains constant during the
Ricci-Hamilton evolution, namely,

\begin{eqnarray}
vol \left( S^2(\eta)\times {\rm I\!R}^1 \right)=
\frac{\int_{S_{(2)}^2(\eta)}w_k(\eta;\alpha) d{\sigma}^k-
\int_{S_{(1)}^2(\eta)}w_k(\eta;\alpha) d{\sigma}^k
}{<R>-<R>_{(\alpha=3)} }.
\end{eqnarray}
\vskip 0.5 cm

\noindent Notice that
by introducing, as above, a new variable
\begin{eqnarray}
{\rm dist}_{\eta}(x,y_{(i)})=\exp[-\tau (\eta)],
\end{eqnarray}

\noindent with
\begin{eqnarray}
\tau (\eta)=\frac{\tau}{[1-\frac{2}{3}\eta]},
\end{eqnarray}

\noindent and by rescaling the unit two-sphere metric
$h_{ij}$ according to

\begin{eqnarray}
h_{ij}(\eta)=[1-\frac{2}{3}\eta] h_{ij},
\end{eqnarray}

\noindent the above analysis of the Ricci-Hamilton flow (\ref{ham2})
provides on the neck the metric

\be
\label{cymetric}
{\tilde g}(x;\eta)=d{\tau}^2(\eta)+ h_{ij}(\eta)d{\theta}^id{\theta}^j
+{\cal O}(e^{-2\tau (\eta)}),
\ee

\noindent and  on $S^3_{(i)}$ the standard round metric.
\vskip 0.5 cm

Notice also that (\ref{cymetric}) goes cylindrical exponentially fast
in $\eta$. As $\eta$ increases $0\leq\eta < \frac{3}{2}$, the neck
becomes longer and longer while getting thinner and thinner. In the
limit, $\eta\to\frac{3}{2}$, we get
\begin{eqnarray}
(S^3_{(1)}\sharp S^3_{(2)}, {\tilde g})
\to S^3_{(1)}\coprod S^3_{(2)},
\end{eqnarray}
where the three-spheres $S^3_{(i)}$ carry the round metric of volume
$1$, while the smooth joining regions shrink exponentially fast around
the points $y_{(i)}$, $i=1,2$.
\vskip  1 cm

\subsection{Critical surfaces and topological crossover}

Let $g_o$ be a metric on $\Sigma$ with positive Ricci curvature,
\begin{eqnarray}
Ric(g_o)>0
\end{eqnarray}
and with volume $vol(\Sigma,g_o)=2$.\par By means of $g_o$ and the
metric ${\tilde g}$ introduced in the previous section, we can
construct on $\Sigma$ a smooth one-parameter, ($0\leq t\leq 1$),
family of metrics $g_t$, with $g_{t=0}=g_o$ and $g_{t=1}= {\tilde g}$
by setting
\begin{eqnarray}
& & g_t \equiv (1-t)g_o+t{\tilde g},\nonumber\\ & & 0\leq t \leq 1.
\end{eqnarray}
\vskip 0.5 cm
According to Hamilton's theorem \cite{Ham}, there is a right-open
neighborhood of $t=0$ such that all metrics $g_t$ in this neighborhood
are in $SU(2)$-Basins and are attracted, under the action of the
Ricci-Hamilton flow (\ref{ham}), towards the round metrics on $S^3$
with volumes $v_t$, (since the volume of $g_t$ changes with $t$, we
have a family of fixed points, namely, round three-spheres
parameterized by the corresponding volumes $v_t$ which are kept
constant under the Ricci-Hamilton flow; thus $v_{t=0}=2$).\par
\vskip 0.5 cm
On the other hand, according to the remarks above, we have an open
neighborhood of $t=1$ such that all metrics $g_t$ in this neighborhood
go {\it singular} under the Ricci-Hamilton flow.  Indeed, for $t=1$,
the Ricci-Hamilton flow of ${\tilde g}$ fixes the round $S^3$ factors
while the joining tube, $S^2\times {\rm I\!R}^1$, is driven towards a
curvature singularity. By continuity, this behavior extends to a
left-open neighborhood of $t=1$, and a three-sphere in this
neighborhood splits apart into two round spheres.
\vskip 0.5 cm
It follows that there is a neighborhood of the ${\tilde g}$ metric
such that some of the metrics in this neighborhood are driven towards
attractive $SU(2)$-Basins, while others are driven towards a {\it
singular} $(S^3_{(1)}\sharp S^3_{(2)})$-Basin.\par
\vskip 0.5 cm
The set of metrics driven towards this basin defines a {\it critical
surface}. Below we shall define a critical fixed point. Since the
Ricci-Hamilton flow preserves the metric ${\tilde g}$, {\it up to a
trivial rescaling} of the neck in $S^3_{(1)}\sharp S^3_{(2)}$,
(rigorously speaking this is true only up to exponentially small
correction terms which arise in the regions joining the three-spheres
with the $S^2\times {\rm I\!R}^1$ neck), we can characterize the {\it
critical fixed point} for (\ref{ham}), in the example considered, as
the metric on $S^3_{(1)}\sharp S^3_{(2)}$ given by
\begin{eqnarray}
{\cal G}_{crit}\equiv \{{\tilde g}(x;\eta)| \,\,\eta\to\frac{3}{2}\},
\end{eqnarray}
where ${\tilde g}(x;\eta)$ is the one parameter $\eta$-family of
metric solution of the Ricci-Hamilton initial value problem
(\ref{ham2}).  This is a consistent characterization of a critical
fixed point for (\ref{ham}), since (\ref{ham}) fixes $(S^3_{(1)}\sharp
S^3_{(2)}, {\cal G}_{crit})$, and as required for (\ref{crit}), the
correlation length associated with $(S^3_{(1)}\sharp S^3_{(2)}, {\cal
G}_{crit})$ is $L=\infty$.  This is so since, roughly speaking, in
order to describe $(S^3_{(1)}\sharp S^3_{(2)}, {\cal G}_{crit})$ we
fix any representative $(S^3_{(1)}\sharp S^3_{(2)})$ which is
characterized by a correlation length $L_o\simeq$ {\it the length of
the neck}, and then rescale with the Ricci-Hamilton flow (\ref{ham2})
from $L_o$ to $L=\infty$.\par
\vskip 0.5 cm
Notice however that in this particular case, the critical point is not
to be related to a phase transition. As mentioned before, we are here
in presence of significant finite-size effects which are concerned
with a dimensional crossover and they show up, as usual, as an {\it
effective reduction of dimensionality} \cite{Cardy}. In this case, the
three-dimensional neck of $(S^3_{(1)}\sharp S^3_{(2)})$ goes
one-dimensional. This crossover to (quasi-)one-dimensional behavior is
not accompanied by a singular behavior in thermodynamical quantities,
such as correlation functions, but nonetheless an anomalous behavior
is present. Indeed, geometrically speaking, the critical fixed point
$(S^3_{(1)}\sharp S^3_{(2)}, {\cal G}_{crit})$, and the corresponding
{\it critical surface}, separates two stable phases under the
renormalization generated by (\ref{ham}). One is given by the
manifolds eventually evolving towards the round three-sphere of volume
$2$. The other is generated by those manifolds which pinch off and
eventually evolve towards two round three-spheres, each of volume $1$.
The {\it pinching off} through thinner and thinner necking is
necessary for such a topological crossover.\par
\vskip 0.5 cm
 From a physical point of view, and as we shall see in section 6, the
fixed point described above separates two possible different closed
FLRW regimes. One with closed spatial sections which, at some
particular instant, are a three-sphere of volume $2$, while in the
other regime we have {\it two} distinct closed FLRW universes having
spatial sections (at a given instant) of volume $1$.  We may have also
many different necks corresponding to a regime whereby the spatial
section $\Sigma$ yields for many closed FLRW universes.

\vskip 0.5 cm

It is clear that the above explicit construction of a critical fixed
point for (\ref{ham}) can in principle be generalized to more general
situations. The strategy is to take two or more trivial fixed points,
such as those associated with the $SU(2)$ and $H(3)$-Basins, and
connect them through the degenerate basins (such as $SO(3)\times {\rm I\!
R}^1$, as in the above example, or through the $H(2)\times {\rm I\!
R}^1$-Basin, etc.).  In this connection notice that the connected sum
mechanism yielding for the $(S^3_{(1)}\sharp S^3_{(2)})$-critical
fixed point, can be generalized to an operation of joining two (or
more) manifolds (corresponding to stable attractors) along tubular
neighborhoods of surfaces (rather than points, as in the case for the
standard connected sum).
\vskip 0.5 cm
A particularly interesting connecting geometry would be $H(2)\times
{\rm I\!R}^1$, (by compactifying the hyperbolic plane in a closed
surface). In this latter case the scale of the hyperbolic geometry
goes to infinity under the Ricci-Hamilton flow ({\it pancake
degeneracy}
\cite{Jackson}). Finite size effects are again at work, but this
time the effective reduction of dimensionality is more interesting
than the one in the $(S^3_{(1)}\sharp S^3_{(2)}, {\cal G}_{crit} )$
case. Indeed, two out of three dimensions are infinite, and there will
be a crossover to a critical behavior with critical exponents
characteristic of a two-dimensional system.
\vskip 0.5 cm
One can consider the analysis presented above as a physically
non-trivial application of Hamilton-Thurston's geometrization program.
In some of its aspects it is rather conjectural and speculative, but
in our opinion, it is quite intriguing that motivations coming from
geometry and a physical problem, like the one addressed here, namely,
the construction of cosmological models out of a local gravitational
theory, go hand in hand in such a way.
\bigskip

\section{Linearized RG flow}

The relative slopes of $<f>_{\epsilon_m}(g_1)$ and
$<f>_{\epsilon_m}(g_2)$, with $f$ given by equation (\ref{sources}),
as $m\rightarrow\infty$, and for $g_1$ in a neighborhood of $g_2$, are
of some relevance to our discussion. In a standard RG analysis such
relative slopes are related to critical exponents.  Given the blocking
procedure for $<f>_{\epsilon_m}(g_1)$ as $m
\rightarrow\infty$, yielding for a $g_1$ ``renormalized"
according to the Ricci- Hamilton flow (\ref{ham}), one can sensibly
ask what happens if $g_1$ is slightly perturbed, namely, if we replace
it by
\begin{equation}
g_{ab} \rightarrow g_{ab} +\delta K_{ab},
\end{equation}
where $K_{ab}$ is a symmetric bilinear form (a choice of the symbol is
quite intentional, since later the above consideration will be applied
to the second fundamental form).  It can be shown that if $g_1$ is
scaled, according to the Ricci-Hamilton flow (\ref{ham}), then
$K_{ab}$ gets renormalized according to the linearized Ricci-Hamilton
flow, namely ($\eta$ in the brackets suppressed),
\begin{eqnarray}
\frac {\partial}{\partial \eta} K_{ab}&  = & \frac {2}{3} <R> K_{ab}
+ \frac {2}{3} g_{ab} [ \frac {1}{2} <R g^{ab} K_{ab}> - \frac {1}{2}
<R> <g^{ab} K_{ab}> - \nonumber \\ & & < R^{ab} K_{ab} >] - \Delta_L
K_{ab} + 2 [ div^* (div ( K -
\frac {1}{2} (Tr\, K) \, g ))]_{ab},
  \label{lflow}
\end{eqnarray}
with the initial data $K_{ab}(\eta =0) = K_{ab}$, where $K \in S^2
\Sigma $ is a given symmetric bilinear form, $\Delta_L$ is the
Lichnerowicz-DeRham Laplacian on bilinear forms
\be
\Delta_LK_{ab}\equiv -\nabla^s\nabla_sK_{ab}+R_{as}K^s_{\,b}+R_{bs}
K^s_{\,a}-2R_{asbt}K^{st},
\ee
and the operators $\Delta_L, \, div^*, \, div $ and $Tr$ are
considered with respect to the flow of metric $(g, \eta ) \rightarrow
g(\eta) $, solution of (\ref{ham}). The $div$ (here, minus the usual
divergence) is the divergence operator on $S^2 \Sigma$, $div^*$ is the
$L^2$- adjoint of $div$, acting from the space of vector fields on
$\Sigma$ to $S^2 \Sigma$ (it can be identified with ${{1}\over{2}} [
{\rm Lie\,derivative} ]$ of the metric tensor along a vector field).

Note that $K( \eta )$ solution of the linear (weakly) parabolic
initial value problem (\ref{lflow}) always exists and is unique
\cite{Ham},
and represents an infinitesimal deformation of metrics connecting the
two neighbouring flows of metrics, $ g( \eta ) $ and $ g' ( \eta )$,
(obtained as solutions of problem (\ref{ham}) with initial data $ g(
\eta =0) = g $ and $ g' ( \eta =0 ) = g(\eta = 0) + \epsilon K(\eta =
0) + {\cal O} (\epsilon^2) $, respectively).  For what concerns the
structure of this solution, one can verify that corresponding to the
``trivial" initial datum $K(\eta=0)=L_Xg$ (where $X:\Sigma\rightarrow
T\Sigma$ is a smooth vector field on $\Sigma$), the solution of
(\ref{lflow}) is provided by
\begin{equation}
K_{ab}(\eta)=L_X g_{ab}(\eta).
\end{equation}
This property expresses the $Diff(\Sigma)$ equivariance of the
Ricci-Hamilton flow. (Notice that $X$ is $\eta$-independent).

The above fact follows by noticing that along the trajectories of the
flow $(\eta, g)\rightarrow g(\eta)$, solution of (\ref{ham}), we have
\begin{equation}\label{3.4}
\frac{\partial}{\partial\eta}L_Xg_{ab}(\eta)=L_X[\frac{\partial}{
\partial\eta}g_{ab}(\eta)]=\frac{2}{3}<R(\eta)>_{\eta}
L_Xg_{ab}(\eta)-2L_XR_{ab}(\eta).
\end{equation}
But the $Diff(\Sigma)$-equivariance of the Ricci tensor, i.e. the fact
that $Ric(\varphi^*g)=\varphi^*Ric(g)$ for any smooth diffeomorphism
$\varphi:\Sigma\rightarrow \Sigma$, implies that
\begin{equation}\label{3.5}
L_XR_{ab}=D\,Ric(g)\cdot L_Xg_{ab},
\end{equation}
where $D\,Ric(g)K$ is the formal linearization of $Ric(g)$, around
$g$, in the direction $K$:
\begin{eqnarray}\label{3.6}
D\,Ric(g)\cdot K& \equiv & \frac{d}{dt}[Ric(g+tK)]_{t=0} \nonumber\\ &
= & \frac{1}{2}\Delta_L K- div^*[div(K-\frac{1}{2}(Tr\,K)g)].
\end{eqnarray}
Upon introducing (\ref{3.5}) in (\ref{3.4}) we get
\begin{equation}
\frac{\partial}{\partial\eta}L_Xg_{ab}(\eta)=\frac{2}{3}<R(\eta)>_{
\eta}
L_Xg_{ab}(\eta)-2D\,Ric(g(\eta))\cdot L_Xg_{ab}(\eta).
\end{equation}
One can check that the right hand side of the above expression
coincides with the right hand side of (\ref{lflow}), when this latter
is evaluated for $K_{ab}(\eta)=L_Xg_{ab}(\eta)$. Hence
$L_Xg_{ab}(\eta)$ solves the partial differential equation
(\ref{lflow}) and, since for $\eta=0,\,K_{ab}=L_Xg_{ab}$, the
uniqueness of any solution of the initial value problem (\ref{lflow})
implies that $K_{ab}(\eta)= L_Xg_{ab}(\eta)$, whenever
$K_{ab}(\eta=0)=L_Xg_{ab}$, as stated.

Moreover, if $K(\eta)$ is a solution of (\ref{lflow}), with initial
datum $K(\eta=0)=K$, then the space average of $Tr\,K(\eta)$ over
$(\Sigma, g(\eta))$ is preserved along the flow $(\eta, g)\rightarrow
g(\eta)$, namely,
\begin{equation}
<Tr\,K(\eta)>_{\eta}=<Tr\,K>_o,\,\,\,\,\,0\leq\eta<\infty.
\end{equation}
This property of the solutions of (\ref{lflow}) is an immediate
consequence of the volume-preserving character of the Ricci-Hamilton
flow.
\bigskip

Finally, another relevant property of the initial value problem
(\ref{lflow}) can be stated as follows. If $(\eta, K_{ab})\rightarrow
K_{ab}(\eta)$ is the flow solution of (\ref{lflow}), with initial
datum $K_{ab}(\eta=0)=K_{ab}$, then it can always be written as
\cite{Lott}
\begin{equation}\label{3.7}
K_{ab}(\eta)=\hat{K}_{ab}(\eta)+L_{v(\eta)}g_{ab}(\eta),
\end{equation}
where the bilinear form $\hat{K}_{ab}(\eta)$ and the $\eta- $dependent
vector field $v(\eta)$, respectively, are the solutions of the initial
value problems:
\begin{eqnarray}\label{3.8}
 & &
\frac{\partial}{\partial\eta}\hat{K}_{ab}=\frac{2}{3}<R>\hat{K}_{ab}
+\frac{2}{3}g_{ab}[\frac{1}{2}<Rg^{ab}\hat{K}_{ab}>-
\frac{1}{2}<R><g^{ab}
\hat{K}_{ab}>-\nonumber\\
 & & <R^{ab}\hat{K}_{ab}>]-\Delta_L\hat{K}_{ab},\nonumber\\ & &
\hat{K}_{ab}(\eta=0)=K_{ab},
\end{eqnarray}
and
\begin{equation}\label{3.9}
\frac{\partial}{\partial\eta}v_a(\eta)=-\nabla^c(\hat{K}_{ca}-
\frac{1}{2}
\hat{K}^{rs}g_{rs}g_{ca}),\,\,\,\,\,v(\eta=0)=0.
\end{equation}
\bigskip

To summarize, as $\eta\rightarrow\infty$, $K_{ab}(\eta)$ may either
approach a Lie derivative term, such as $L_{v(\eta)} g_{ab}(\eta)$, or
a non-trivial deformation $\hat{K}_{ab}(\eta)$
\cite{Lott}. The non-trivial deformation is present only if the
corresponding Ricci-Hamilton flow for $g_{ab}(\eta)$ approach an
Einstein metric on $\Sigma$ which is not isolated.  In such a case,
(e.g. flat tori), there is a finite dimensional set of such Einstein
metrics, and the non-trivial $\hat{K}_{ab}$ simply are the
infinitesimal deformations connecting one Einstein metric $\bar{g}_1$
in $\Sigma$ and the infinitesimally non-equivalent one.  Also in this
case the Lie derivative term may be present.  What it represents is a
reparameterization of the metric $\bar{g}_1$ (under the action of the
infinitesimal group of diffeomorphisms generated by $v$ (see equation
(\ref{3.7})), (``gauge artifact").  This latter Lie derivative term is
the only surviving term when $\bar{g}$ is isolated (like e.g. in the
case of the round three-sphere.  As is known, the round metric
$\bar{g}$ on the three-sphere $S^3$ is isolated, in the sense that
there are not volume-preserving infinitesimal deformations of
$\bar{g}$ mapping it to another inequivalent constant curvature metric
$\bar{g}'$.)
\vskip 0.5 cm
To summarize, the flow (\ref{3.8}), (\ref{3.9}), tells us how to
renormalize the second fundamental form in such a way that the
blocking prescription $<f>_{\epsilon_m}$ $\to$ $<f>_{\epsilon_{m+1}}$
works both for the initial metric $g$ as well as for the {\it
perturbed} metric $g+\delta K_{ab}$.
\bigskip
\subsection{Scaling and critical exponents}

 From the characterization of critical fixed points $g^{\star}_{ab}$
for (\ref{ham}), (see section 4.2), we can get information on the
critical exponent characterizing critical behavior of the metrics
nearby $g^{\star}_{ab}$.  This is the content of this section. In
particular, we shall discuss the critical exponents related to the
critical fixed point $(S^3_{(1)}\sharp S^3_{(2)}, {\cal G}_{crit})$.
Even if this point is not a thermodynamically interesting critical
point, it exhibits many of the general features of the more
interesting type of singularities.
\vskip 0.5 cm

In order to characterize these critical exponents we can use the
linearized Ricci-Hamilton flow associated with the one-parameter
family of metrics $(S^3_{(1)}\sharp S^3_{(2)}, {\cal G}_{crit})$. In
the following we shall however proceed more directly and examine the
properties of the two-point correlation function associated with
$(S^3_{(1)}\sharp S^3_{(2)}, {\cal G}_{crit})$ and the probability law
of relevance to our analysis.
\vskip 0.5 cm
Let $y_i\in S^3_{(i)}$, $i=1,2$, be the two points in
$(S^3_{(1)}\sharp S^3_{(2)}, {\cal G}_{crit})$ around which the round
metrics of $S^3_{(i)}$ have been blown up. Let $f$ denote a
non-negative scalar field on $(S^3_{(1)}\sharp S^3_{(2)})$,
distributed according to the probability law formally defined by
\be
\label{functional}
dP\equiv \frac{\exp[-H(f)]\prod_xdf(x)}{\int\exp[-H(f)]\prod_xdf(x)},
\ee
where the functional integration is over the space of fields
$f\colon(S^3_{(1)}\sharp S^3_{(2)})\to {\rm I\!R}$, equipped with the
$L^2$-inner product
\be
\label{inner}
(f|f')\equiv \int_{(S^3_{(1)}\sharp S^3_{(2)})} f(x)f'(x)d\mu_{g}(x),
\ee
and where
\be
\label{acca}
H(f)=\int_{(S^3_{(1)}\sharp S^3_{(2)})} f(x)d\mu_{g}(x).
\ee

\noindent (We could consider f as related to the
matter fields, as in $f=\alpha\rho+\alpha^iJ_i$, but the following
analysis is quite independent from a particular meaning of $f$).
Notice also that the measure $d\mu_{g}(x)$ in the above formulae is
the riemannian measure on $(S^3_{(1)}\sharp S^3_{(2)})$ associated
with the metric ${\tilde g}(x)$ defined by (\ref{symmetric}).
\vskip 0.5 cm
Let us now concentrate on the behavior of (\ref{functional}) when the
neck of $(S^3_{(1)}\sharp S^3_{(2)})$ gets thinner and longer under
the action of (\ref{ham}).

It can then be checked that along the Ricci-Hamilton flow
(\ref{jimandjack}) associated with ${\tilde g}|_{{\rm neck}}$, both
the $L^2$-inner product (\ref{inner}) and the Gibbs factor
$\exp[-H(f)]$ corresponding to (\ref{acca}), are invariant.  Thus it
follows that the probability measure (\ref{functional}) is invariant
under (\ref{ham}), and the correlation function defined by
\begin{eqnarray}
E_{dP}[f(y_1)f(y_2)],
\end{eqnarray}
where $E_{dP}$ denotes the expectation with respect to $dP$, is well
defined over the {\it critical} fixed point ${\cal G}_{crit}$.
\vskip 0.5 cm
In section 4.1 we have interpreted the Ricci-Hamilton flow (\ref{ham})
as the RG flow in a finite geometry, characterized by the length scale
$L$, which is large compared to the microscopic scale (in particular,
it is much larger than the radius of the geodesic ball coverings used
to discretize the theory).  The correlation function depends on such a
dimensional parameter.  If
\begin{eqnarray}
L(y_1,y_2)\equiv \int_{y_1}^{y_2}\sqrt{D}dz
\end{eqnarray}
denotes the distance between the points $y_1$ and $y_2$ along the
cylindrical neck of $(S^3_{(1)}\sharp S^3_{(2)})$, and if this
distance is large (as compared with the radius of geodesic ball
coverings), then the {\it correlation length} $\xi$ associated with
the two-point (connected) correlation function can be read off from
\begin{eqnarray}
\xi\simeq_{L(y_1,y_2)\gg 1}\frac{-L(y_1,y_2)}{\ln
E_{dP}(f(y_1)f(y_2))_{conn}}.
\end{eqnarray}

\noindent Since the correlation function remains invariant under the
Ricci-Hamilton deformation of the cylindrical neck, we get that along
(\ref{jimandjack}), $\xi /L(y_1,y_2)$ remains constant which implies
that on $(S^3_{(1)}\sharp S^3_{(2)}, {\cal G}_{crit})$ the correlation
length $\xi$ behaves as

\begin{eqnarray}
\xi\simeq\sqrt{D_0}E_0[E_0-\frac{2}{3}\eta]^{-\1}.
\end{eqnarray}
\vskip 0.5 cm
According to standard usage, we can define the {\it critical exponent}
$\nu$ associated with the correlation length of a finite size system
(with typical size $L$) by the condition

\be
\label{Jug}
\frac{\partial}{\partial p}{\xi}(L,p)|_{p=p_c}
\simeq L^{1+\frac{1}{\nu}},
\ee
\vskip 0.5 cm
\noindent where $p$ is a parameter driving the system to
criticality, and $p_c$ is its corresponding critical value.  In our
case, it is natural to set $p=\eta$, with $p_c=\eta_c=\frac{3}{2}$.
This immediately yields for the critical exponent $\nu$ the value
\begin{eqnarray}
\nu = 1.
\end{eqnarray}
\vskip  1 cm
A similar computation for the critical exponent associated with the
correlation length can be carried out when the connecting geometry is
$H^2\times{\rm I\!R}^1$. This takes place when two riemannian manifolds
$M_1$ and $M_2$, which are supposed to evolve nicely under the
Ricci-Hamilton flow, are connected through a tubular neighborhood of a
surface $S_h$ of genus $h$, {\it viz.}, $\sigma=M_1\sharp_{S_h}M_2$.
\vskip 0.5 cm
In this case, the metric on the neck can be written as
\cite{Jackson}
\begin{eqnarray}
g_{neck}=Dg_{{\rm I\!R}^1}+Eg_{H^2},
\end{eqnarray}

\noindent where $g_{H^2}$ is the metric on the hyperbolic
plane. The Ricci-Hamilton flow equations take a form similar to the
$S^2\times{\rm I\!R}^1$ case, up to a minus (important!)  sign, namely,

\begin{eqnarray}
\frac{d}{d\eta}E &= & \frac{2}{3},\nonumber\\
\frac{d}{d\eta}D &= & -\frac{4}{3}\left( \frac{D}{E} \right),
\end{eqnarray}

\noindent which upon integration provide

\begin{eqnarray}
E &= & E_0+\frac{2}{3}\eta,\nonumber\\ D &= &
\frac{D_0E_0^2}{[E_0+(2/3)\eta]^2},
\label{jjimandjack}
\end{eqnarray}

\noindent where $E_0^2=E^2(\eta=0)$ is the initial scale of the
hyperbolic geometry, whereas $D_0=D(\eta=0)$ is the scale on ${\rm I\!
R}^1$.
\vskip 0.5 cm
The scale $E$ of the hyperbolic geometry increases linearly with
$\eta$, while the scale factor $D$ on the line ${\rm I\!R}^1$ shrinks.
It can be checked \cite{Jackson}, that corresponding to this scale
dynamics, the curvature decays according to
$||Ric||=\sqrt{2}/(E_0+\frac{2}{3}\eta)$, and we get in the limit
$\eta\to\infty$ a {\it pancake} degeneracy.
\vskip 0.5 cm
The relevant correlation function is now
$E_{dP}(f(y_1)f(y_2))_{conn}$, with $y_1$ and $y_2$ fixed points in
the $H^2$ factor, (i.e., on the surface $S_h$). Again, owing to the
symmetries of the geometry involved, it immediately follows that the
ratio between the correlation length $\xi$ and the distance
$L(y_1,y_2)$ must remain constant under the Ricci-Hamilton flow. This
implies that the correlation length behaves as
\begin{eqnarray}
\xi\simeq \sqrt{E_0+\frac{2}{3}\eta},
\end{eqnarray}
to which corresponds again a critical exponent $\nu=1$, (in this case
one cannot apply (\ref{Jug}) since the system is actually going to an
infinite size).
\vskip 0.5 cm
A striking feature of these topological crossover phenomena,
associated with the renormalization of the cosmological matter
distribution, is that their pattern resembles the linear sheet-like
%kam
(or sponge-like) structure in the distribution of galaxies on large
scales (e.g. \cite{Geller}).  It is evident that if,
$(\Sigma,g,K,\rho, {\bf J})$, the initial data set for the real
universe (see the next section), is close to {\it criticality}, in the
sense discussed in the previous section, then the corresponding
averaged model exhibits a tendency to topological crossover in various
regions (the ones where the inhomogeneities are larger).
Filament-like and sheet-like structures would emerge, and the overall
situation would be the one where such structures appear together with
regions of high homogeneity and isotropy, in some sort of hierarchy.
This situation is akin to that of a ferromagnet nearby its critical
temperature, whereby we have islands of spins up and down in some sort
of nested pattern.
%kam (til end of this section)
Even if there is a tendency to homogeneity at very large scales, the
picture just sketched, of the ``hierarchy of structures'', might be
qualitatively valid in a good part of the universe; indeed recent
observational data seem to suggest the existence of still larger and
larger structures (e.g. \cite{Clowes}). Notice that this picture bears
some resemblance to the ``cascade of fluctuations'' in critical
phenomena \cite{cascade}. Droplet fluctuations nucleated at the
lattice scale in the critical state can grow to the size of the
correlation length where the details of the lattice structure become
lost and the scale invariant distributions of the large ``droplets''
are universal.

A possible (though not yet clarified) connection of this whole picture
with the self-organized criticality (SOC) \cite{Bak} can be envisaged,
whereby the problem of structure formation, in terms of growth
phenomena, could be tackled in the framework of avalanche activity
used in SOC (cf. \cite{Lee}). It would be particularly interesting to
estimate the scales in the universe, where it is necessarily critical
and trapped into self-organized (critical) states (similar suggestion
was recently posed in \cite{Rosu}).

\vskip 1 cm

\section{Effective cosmological models}

The results of the previous sections have interesting consequences
when applied in a cosmological setting. \par
\vskip 0.5 cm
Let us assume that at the scale over which General Relativity is
experimentally verified, a cosmological model of our universe is
provided by evolving a set of consistent initial data $(\Sigma, g,K,
\rho,{\bf J})$, according to the evolutive part of Einstein's
equations.

The data $(\Sigma, g,K, \rho,{\bf J})$ describing the interaction
between the actual distribution of sources and the inhomogeneous
geometry of the physical space, $(\Sigma, g,K)$, are required to
satisfy the Hamiltonian and the divergence constraints. Respectively,
\begin{equation}
\label{constraint}
8\pi G \rho= R(g) -K^{ab}K_{ab}+ k^2,
\end{equation}
\be
\label{constraint2}
\nabla^iK_{ih}-\nabla_hk=16\pi G\,J_h,
\ee
where, $k\equiv K^a{}_a$, (see also section 4, where they were written
in terms of the three-metric $g_{ab}$ and its associated conjugate
momentum $\pi_{ab}$).
\vskip 0.5 cm
According to the blocking and renormalization procedure, discussed at
length in this paper, we can implement a coarse-graining
transformation on this actual data set by suitably renormalizing the
metric and the second fundamental form, while retaining the functional
form and the validity of the constraints.
\vskip 0.5 cm
The metric is renormalized according to the Ricci-Hamilton flow
(\ref{ham}),
\begin{eqnarray}
\left\{ \begin{array}{ll}
\frac{\partial
g_{ab}(\eta)}{\partial\eta} &= \frac{2}{3}<R(\eta)>
g_{ab}(\eta)-2R_{ab}(\eta)\nonumber\\ g_{ab}(\eta=0)&=g_{ab}.
\end{array}\right.
\end{eqnarray}
\bigskip

In order to renormalize the second fundamental form, let us rewrite it
in terms of the {\it deformation tensor} $H_{ab}$ defined by
\be
H_{ab}\equiv K_{ab}+L_{\alpha}g_{ab},
\label{deftensor}
\ee
where $\vec{\alpha}$ is the shift vector field providing the
three-velocity of chosen instantaneous observers on $\Sigma$, with
respect to which we are implementing the renormalization procedure. We
renormalize $K_{ab}$ by rescaling this deformation tensor $H_{ab}$
according to (\ref{lflow}), {\it viz.},

\begin{eqnarray}
\frac {\partial}{\partial \eta} H_{ab}&  = & \frac {2}{3} <R> H_{ab}
+ \frac {2}{3} g_{ab} [ \frac {1}{2} <R g^{ab} H_{ab}> - \frac {1}{2}
<R> <g^{ab} H_{ab}> - \nonumber \\ & & < R^{ab} H_{ab} >] - \Delta_L
H_{ab} + 2 [ div^* (div ( H -
\frac {1}{2} (Tr\, H) \, g ))]_{ab},
\end{eqnarray}
\vskip 0.5 cm
\noindent with the initial data
$H_{ab}(\eta =0) = (K_{ab}+L_{\alpha}g_{ab})_{\eta=0}$.  Notice that
we renormalize the deformation tensor $H_{ab}$, rather than $K_{ab}$
directly, because in this way we can get rid of the possible
$Diff$-induced shear which may develop in $\lim_{\eta\to
\infty}K_{ab}(\eta)$. Indeed,
according to equation (\ref{3.7}), as $\eta\to\infty$ the solution of
this initial value problem, $K_{ab}(\eta)$, approaches a non-trivial
deformation $\lim_{\eta\to\infty}{\hat K_{ab}}(\eta)$ plus a
Lie-derivative term, $\lim_{\eta\to\infty}L_{v(\eta)}g_{ab}(\eta)$.
The former is present only if the Ricci-deformed metric ${\bar
g}_{ab}=\lim_{\eta\to\infty}g_{ab}(\eta)$ is not isolated.
\vskip 0.5 cm
 Since at this stage we are mainly interested in FLRW space-times, let
us assume that ${\bar g}_{ab}$ is isolated, while in order to take
care of the $Diff$-induced shear,
$\lim_{\eta\to\infty}L_{v(\eta)}g_{ab}(\eta)$, we can choose the shift
vector field $\alpha^i$ in such a way that
$\lim_{\eta\to\infty}L_{v(\eta)}g_{ab}(\eta)$ is compensated by
$L_{\alpha}g_{ab}$. Since $L_{\alpha}g_{ab}$ is a trivial datum for
(\ref{lflow}), it is thus sufficient to choose

\be
\label{velocity}
\alpha^i= \lim_{\eta\to\infty}v(\eta).
\ee

Notice that $\alpha^i$ is the three-velocity vector of the chosen
instantaneous observers on $\Sigma$, thus (\ref{velocity}) provides a
map identifying corresponding points between the initial {\it actual}
manifold $(\Sigma,g,K,\rho,{\bf J})$ and its renormalized counterpart
$\lim_{\eta\to\infty}(\Sigma, g, K, \rho, {\bf J})(\eta)$.
\vskip 0.5 cm
In this section, and mainly for actual computational purposes, we
assume that the original inhomogeneous initial data set is such that
the Ricci-Hamilton flow is global. As already mentioned, we are
interested in connecting an inhomogeneous cosmological space-time to
its corresponding FLRW model.  This is the case, in particular, if we
assume that the original manifold $(\Sigma,g)$ has a positive Ricci
tensor, (this case is obviously quite similar to the analysis in
\cite{Mauro}, there are however important differences that we are
going to emphasize). Or more generally, if we assume that the original
manifold $(\Sigma,g)$ is in the $SU(2)$-Basin of attraction or {\it
nearby} the critical point ${\sharp}_{i}S^3_{(i)}$, with
$i=1,2,\ldots$, yielding for a manifold $(\Sigma,g)$, nucleating under
the Ricci-Hamilton flow (\ref{ham2}), (extended to many connected
sums), to disjoint three-spheres $S^3_{(i)}$.
\vskip 0.5 cm
Given this setting, we see that due to the properties of the
Ricci-Hamilton flow we have $\lim_{\eta\rightarrow
\infty}K_{ab}(\eta)=\frac{1}{3}\bar{k}\cdot\bar{g}_{ab}$. The given $K_{ab}$
is deformed by gradual elimination of its shear $K_{ab}-
\frac{1}{3}kg_{ab}$
and the original (position dependent) rate of volume expansion $k$ is
being replaced with its corresponding average value.
\vskip 0.5 cm
Since the constraints, (\ref{constraint}) and (\ref{constraint2}), are
required to hold at each step of the renormalization procedure, we get
\begin{equation}
\label{etaconstraint}
8\pi G(\eta) \rho(\eta)= R(g(\eta)) -K^{ab}(\eta)K_{ab}(\eta)+
k^2(\eta),
\end{equation}
\be
\label{etaconstraint2}
\nabla(\eta)^iK_{ih}(\eta)-\nabla(\eta)_hk=16\pi G(\eta)\,J_h(\eta),
\ee
where we have explicitly introduced a possible $\eta$-dependence into
the gravitational coupling $G$.

\vskip 0.5 cm
Let us now explore the consequences of (\ref{etaconstraint}) and
(\ref{etaconstraint2}). From the stated hypothesis on the
Ricci-Hamilton flow it follows immediately that, $K_{ab}(\eta)\to
\frac{1}{3}{\bar k}{\bar g}_{ab}$ as $\eta\to\infty$, thus
(\ref{etaconstraint2}) implies that
\be
\lim_{\eta\to\infty}J_h(\eta)=0.
\ee
\vskip 0.5 cm
In order to analyze (\ref{etaconstraint}), we will make use of a
property of the Ricci-Hamilton flow, namely that the flow $K(\eta)$,
solution of (\ref{lflow}), is such that $\frac{\partial}{
\partial\eta}<k(\eta)>_{\eta}=0$, i.e. the space average of the
trace of the second fundamental form remains constant during the
deformation.  This allows us to write
\begin{equation}
\label{krz1}
<k>_o^2 = {\bar{k}}^2,
\end{equation}
since in the limit, the volume expansion is simply a constant, and
where $<\ldots>_o$ denotes the full average of the enclosed quantity
with respect to the initial metric.
\vskip 0.5 cm
Equation (\ref{krz1}) provides the Hubble constant on the FLRW time
slice associated with the smoothed data corresponding to $(\Sigma, g,
K, \rho, {\bf J})$.
\vskip 0.5 cm
More explicitly, let us write the FLRW metric in the standard form
(units $c=1$)
\be
ds^2= -dt^2+ S^2(t)d\sigma^2,
\ee
where $d\sigma^2$ is the metric of a three-space of constant curvature
and it is time independent. As we are interested in the three-sphere
case, the metric $d\sigma^2$ can be written as
\be
d\sigma^2=d\chi^2+\sin^2\chi(d\theta^2+\sin^2\theta d\phi^2).
\ee
Since the volume $vol(\Sigma,g)$ of the original inhomogeneous
manifold is preserved by the Ricci-Hamilton flow, we can relate the
factor $S^2(t)$, providing the inverse (sectional) curvature of the
FLRW slice $t=t_o$, to $vol(\Sigma, g)$
\be
S^2(t_o)={\left(\frac{vol(\Sigma,g)}{2\pi^2} \right)}^{\frac{2}{3}}.
\ee
Notice in particular that the scalar curvature, towards which
$R(g(\eta))$ evolves under the Ricci-Hamilton flow, is given by
\be
\bar{R}\equiv\lim_{\eta\to\infty}R(g(\eta))
=\frac{3}{S^2}=3 {\left(\frac{vol(\Sigma,g)}{2\pi^2}
\right)}^{-\frac{2}{3}}.
\ee
\vskip 0.5 cm
Having said this, the equation (\ref{etaconstraint}) becomes in the
limit, after extracting a trace free part of $K_{ab}$,
\begin{equation}
8\pi {\bar{G}} \bar{\rho} =\bar{R} +\frac{2}{3}{\bar{k}}^2,
\label{kropa}
\end{equation}
since no residual shear survives, and where we have introduced the
renormalized gravitational coupling
\be
\bar{G}\equiv \lim_{\eta\to\infty} G(\eta).
\ee

\vskip 0.5 cm
On the other hand (\ref{etaconstraint}) gives
\begin{equation}
\frac{2}{3}k^2(\eta) = 8\pi G\rho(\eta) -R(g(\eta))
+\tilde{K}^{ab}(\eta)\tilde{K}_{ab}(\eta),
\end{equation}
where the shear $\tilde{K}_{ab}\equiv K_{ab}- \frac{1}{3}k g_{ab}$ has
been explicitly introduced.
\vskip 0.5 cm
We can rewrite the last equation upon taking the average with respect
to the initial metric, as
\begin{equation}
\frac{2}{3}<k^2>_o =
8\pi G<\rho>_o - <R>_o + <\tilde{K}^{ab} \tilde{K}_{ab}>_o.
\label{palle}
\end{equation}
\vskip 0.5 cm
\noindent By exploiting (\ref{krz1}) and (\ref{palle}) we can get
the Hubble constant of the FLRW model we want to use to describe the
time evolution of the renormalized initial data set
$\lim_{\eta\to\infty}(\Sigma,g,K,\rho, {\bf J})(\eta)$.

\subsection{The renormalized Hubble constant}

Taking into account (\ref{krz1}), and noticing that the Hubble
constant on the FLRW slice corresponding, via the Ricci-Hamilton flow,
to $(\Sigma, g, K)$ is
\be
H^2_o={\left( \frac{1}{S}\frac{dS}{dt} \right)}^2=
\frac{2}{9}{\bar k}^2,
\label{tautological}
\ee

\noindent we get

\be
\label{Hubble}
H^2_o= \frac{8\pi G}{3}<\rho>_o -\frac{1}{3} <R>_o +
\frac{1}{3}<\tilde{K}^{ab} \tilde{K}_{ab}>_o
-\frac{2}{9}(<k^2>_o- <k>_o^2).
\ee
\vskip 0.5 cm
\noindent This is a rather tautological rewriting of
(\ref{tautological}), justified by the fact that in this form $H_o^2$
clearly shows a contribution from the spatial average of the shear. In
particular, if originally (i.e., for $\eta=0$) $k$ is spatially
constant, (\ref{Hubble}) reduces to
\be
H^2_o= \frac{8\pi G}{3}<\rho>_o -\frac{1}{3} <R>_o +
\frac{1}{3}<\tilde{H}^{ab} \tilde{H}_{ab}>_o+
\frac{1}{3} {<{\tilde{L}}_{\alpha}g_{ab}
{\tilde{L}}_{\alpha}g_{cd}g^{ac}g^{bd}>}_o,
\ee
where we have explicitly used the decomposition
$K_{ab}=H_{ab}-L_{\alpha}g_{ab}$, (see equation (\ref{deftensor})),
and
\be
{\tilde{L}}_{\alpha}g_{ab}\equiv \nabla_a\alpha_b+
\nabla_b\alpha_a-\frac{2}{3}g_{ab}\nabla_c{\alpha}^c
\ee
is the conformal Lie derivative of the metric $g_{ab}$ along the
vector field $\alpha$.

\vskip 0.5 cm
\noindent The above expressions for $H_o$ correspond,
through (\ref{kropa}), to the matter density distribution
\begin{equation}
\label{effective1}
8\pi\bar{G}\bar{\rho}=8\pi G<\rho>_o+ \bar{R} - <R>_o+<\tilde{K}^{ab}
\tilde{K}_{ab}>_o -\frac{2}{3}(<k^2>_o- <k>_o^2).
\end{equation}
\vskip 0.5 cm
We have to consider this expression for $\bar{G}\bar{\rho}$ as the
{\it renormalized} effective sources entering into the Friedmann
equation, if we want to describe the real locally inhomogeneous
universe through a corresponding idealized FLRW model. As expected,
the renormalized matter density shows contributions of geometric
origin, either coming from the shear anisotropies or from the local
fluctuations in curvature and volume expansion.
\vskip 0.5 cm
At this stage, a few comments are in order concerning, in particular,
the {\it effective Hubble constant} $H_o$ (\ref{Hubble}).  First, we
wish to emphasize that this is the theoretical expression for the
Hubble constant if {\it one wishes to model the real locally
anisotropic and inhomogeneous universe with a corresponding FLRW one}.
Expression (\ref{Hubble}) clearly shows that apart from the expected
contributions to the expansion rate coming from matter and curvature,
there are two further contributions:\par
\vskip 0.5 cm
\noindent {\it (i)} a negative
contribution coming from the local fluctuations in the expansion rate.
Apparently, this is not a very significant term, since we may choose,
as a convenient initial data set describing the real locally
inhomogeneous universe, a data set supported on a three-manifold
$\Sigma$ of constant {\it extrinsic time}, (i.e., $k$ spatially
constant on $\Sigma$). Actually, the optimal choice of the data set
with respect to which smooth-out the lumpiness in matter and geometry
is a deep and interesting question on which we comment later on, in
section \ref{dataset}.\par
\vskip 0.5 cm
\noindent {\it (ii)}
a {\it positive} contribution coming from the {\it shear term} $
\frac{1}{3}<\tilde{H}^{ab} \tilde{H}_{ab}>_o$, and from
the {\it observer dependent} term
$\frac{1}{3}{<{\tilde{L}}_{\alpha}g_{ab}
{\tilde{L}}_{\alpha}g_{cd}g^{ac}g^{bd}>}_o$ which depends on the
three-velocity $\vec{\alpha}$ of the observers with respect to which
the renormalization is carried out.
\vskip 0.5 cm

These latter terms are the most important non-standard contributions
to $H_o^2$, and have their origin both in the presence of the
gravitational radiation
\be
\frac{1}{3}{<\tilde{H}_{\perp}^{ab} \tilde{H}_{\perp ab}>}_o,
\ee
where $\tilde{H}_{\perp}$ is the divergence-free part of the
deformation tensor ${\tilde{H}}_{ab}$, and in the anisotropies
generated by the motion of matter
\be
\frac{1}{3}{<\tilde{H}_{||}^{ab} \tilde{H}_{|| \, ab}>}_o,
\ee
where $\tilde{H}_{||}$ is the longitudinal part of the deformation
tensor, obtained as a solution to the equation
\be
\nabla^i(\tilde{K}_{||})_{ih}-\frac{2}{3}\nabla_hk=
16\pi G\,J_h +\nabla^i({\tilde{L}}_{\alpha}g_{ih}).
\ee
\vskip 0.5 cm
\noindent Recall that the shift vector field $\vec\alpha$
is connected to the current density $\vec{J}$ by the requirement that
it is {\it chosen} in such a way as to eliminate the longitudinal
shear in $\lim_{\eta\to\infty}K_{ab}(\eta)$. Also, notice that
\be
{<\tilde{H}^{ab} \tilde{H}_{ab}>}_o= {<\tilde{H}_{\perp}^{ab}
\tilde{H}_{\perp ab}>}_o +{<\tilde{H}_{||}^{ab} \tilde{H}_{|| \,
ab}>}_o.
\ee
\vskip 0.5 cm
In practice, the deformation {\it energy} ${<\tilde{H}^{ab}
\tilde{H}_{ab}>}_o$ yields a contribution to the Hubble constant $H_o$
which can be roughly estimated by exploiting the anisotropy
measurements in the cosmic microwave background (CMB) radiation, as
long as the frame used in averaging (i.e., the lapse $\alpha$ and the
shift $\alpha^i$) is, on the average, comoving with the cosmological
fluid. We also require that the original locally inhomogeneous
manifold $(\Sigma,g,K)$ does not differ too much from a standard FLRW
$t=const.$ slice.  In such case one can apply the analysis of
\cite{Maartens} to conclude that, at the present epoch, the ratio
between the deformation shear $\tilde{H}_{ab}$ and expansion is of the
order
\vskip 0.5 cm
\be
\left( \frac{|{\tilde H}_{ab}|}{H} \right)_o< 4\epsilon,
\ee
\vskip 0.5 cm
\noindent where $\epsilon\equiv \max (\epsilon_1, \epsilon_2,
\epsilon_3)$ denotes the upper limit of currently observed anisotropy in the
CMB radiation temperature variation, and where $\epsilon_1$,
$\epsilon_2$, $\epsilon_3$, respectively, denote the dipole,
quadrupole, and octopole temperature anisotropies. On choosing
$\epsilon\simeq 10^{-4}$, as indicated by the recent CMB radiation
anisotropy measurements, one gets that the shear deformation is at
most about $10^{-3}$ of the expansion
\cite{Maartens}.
\vskip 0.5 cm
 Thus, as long as we assume that the original locally inhomogeneous
manifold $(\Sigma,g,K)$ does not differ too much from a standard FLRW
$t=const.$ slice, the contribution to $H_o$ from $<\tilde{H}^{ab}
\tilde{H}_{ab}>_o$ is certainly quite small, and we can reliably write
\vskip 0.5 cm
\be
H^2_o\simeq \frac{8\pi G}{3}<\rho>_o -\frac{1}{3} <R>_o +
\frac{1}{3}{<{\tilde{L}}_{\alpha}g_{ab}
{\tilde{L}}_{\alpha}g_{cd}g^{ac}g^{bd}>}_o.
\ee

\vskip 0.5 cm

The last term involving the shift vector $\vec\alpha$ is a velocity
effect term which is by no means small, at least {\it a priori}. Thus
the renormalized Hubble constant needed, to describe by a model-FLRW a
locally inhomogeneous and anisotropic universe, is {\bf not} provided
by a naive average even if the actual universe, to be smoothed-out in
FLRW-modelling, is not too far from homogeneity and isotropy. Clearly,
we are expecting that in such a case the actual contribution of the
observer-velocity term is rather small, but this is a question which
is difficult to handle. Technically speaking, we would need estimates
on the size of the $Diff$-induced shear generated upon smoothing by
the Ricci-Hamilton flow. This is an issue under current investigation.
We wish also to stress that in this velocity term there is hidden a
non-trivial scale dependence. Indeed, assuming that we wish to model
the actual universe by a FLRW one only up to a certain scale, then the
shift vector $\vec\alpha$ needed to cure the $Diff$-induced shear,
(which is tantamount to saying the proper selection of a frame with
respect to which we are carrying out the partial smoothing), depends
on such a scale, and the larger the scale the larger the contribution.
\vskip 0.5 cm

There is another intriguing possible explanation for a larger than
expected Hubble constant. Indeed, if we take seriously the possibility
that the real universe may be close to the critical phase, as argued
in the previous sections, then the contribution from the shear is not
just of a conceptual value.  For example, the original data set
$(\Sigma,g,K)$ may be near the critical surface associated with the
$\sharp_i S^3_{(i)}$ critical point. In this case we may generate,
upon smoothing, a whole family of disconnected $t=const.$ FLRW slices,
each one with its own Hubble constant $H_o(i)$. Explicitly, let us
assume that the initial data set $(\Sigma,g,K)$ is attracted upon
averaging towards a critical $(S^3_{(1)}\sharp
S^3_{(2)}\sharp\ldots\sharp S^3_{(n)})$, this being the case if the
original metric $g$ exhibits regions of large inhomogeneities.
Assuming, for simplicity, that the rate of volume expansion is
spatially constant, we get that each $S^3_{(i)}$ factor,
$i=1,\ldots,n$, inherits a Hubble constant provided by
\be
\label{Hubbles}
H^2_o(i)= \frac{8\pi G}{3}<\rho>_o -\frac{1}{3} <R>_o +
\frac{1}{3}<\tilde{K}^{ab} \tilde{K}_{ab}>_o.
\ee
These Hubble constants can be quite dominated by the large
anisotropies $<\tilde{K}^{ab} \tilde{K}_{ab}>_o$ of the original
manifold. Indeed, the previous estimate on the smallness of the shear
term is, strictly speaking, valid only in the observable domain
defined by our past light-cone from last scattering to the present
day. Thus, if the $S^3_{(i)}$ factors resulting from the critical
behavior of $(\Sigma,g,K)$ are not smaller than the spatial sections
intercepted by the interior of this light-cone, we would not notice
the contribution from the local shear, (since we would have been
looking at a rather homogeneous and isotropic island), the large
contribution would come from the regions of large inhomogeneities and
anisotropies which, under the Ricci-Hamilton renormalization, undergo
the topological crossover.  Thus, a value of the Hubble constant may
quite well depend on a possible large shear outside our observable
domain.
\vskip 0.5 cm
The behavior just described also shows that the RG procedure developed
{\it preserves} the size of the regions where spatial homogeneity and
isotropy is observed, even if large inhomogeneity may be globally
present. This tells us what are the {\it averaging scales} on which a
FLRW model is a good approximation. More precisely, in order to
determine the averaging scale associated with a FLRW model we want to
use, to describe $(\Sigma,g,K)$, we need to understand more in detail
the (conjectured) Hamilton-Thurston decomposition of $(\Sigma,g,K)$.
The specific example discussed in section
\ref{examplecrit} may serve as an indication.

\subsection{The scale dependence of the matter distribution}

In the above analysis, we also introduced a renormalized gravitational
coupling. In a sense this is superfluous since the three-dimensional
metric $g$ of $\Sigma$ is acting as the running coupling constant, and
we can always reabsorb $G$ into a definition of $g$. Nevertheless, the
use of the renormalized coupling $\bar{G}$ may be helpful if one
wishes to use the standard average of matter $<\rho>_o$ in the
Friedmann equation, rather than the effective matter distribution
$\bar{\rho}$. The explicit expression for $\bar{G}$ can be easily
obtained by setting $\bar{\rho}= <\rho>_o$ in (\ref{effective1}):
\vskip 0.5 cm
\begin{equation}
8\pi\bar{G}=8\pi G+ \frac{\bar{R} - <R>_o+<\tilde{K}^{ab}
\tilde{K}_{ab}>_o -\frac{2}{3}(<k^2>_o-{\bar k}^2)}{<\rho>_o}.
\end{equation}
\vskip 0.5 cm
\noindent Notice however, that it is $G(\eta)\rho(\eta)$
which is inferred from measurements for different scales, and thus the
use of $\bar{G}$ is not particularly remarkable.
\vskip 0.5 cm
In this connection, it is more important to discuss the dependence of
$G(\eta)\rho(\eta)$ as the local scale is varied, namely as $\eta$
increases, (recall that $\eta$ is the logarithmic change of the cutoff
length associated with the geodesic ball coverings).  For simplicity,
we do this only for the case in which no shear is present (${\tilde
K}_{ab}=0$), and the rate of volume expansion is spatially constant
($k=const.$). Under such hypothesis, we get for the scale-dependence
of the average $<G\rho>$
\begin{eqnarray}
\frac{\partial}{\partial\eta}[8\pi<G(\eta)\rho(\eta)>] &=&
\frac{\partial}{\partial\eta}<R(g(\eta))>=\nonumber \\
&= & 2<{\tilde R}^{ik}{\tilde R}_{ik}>+
\frac{1}{3}\left(<R>^2-<R^2>  \right),
\label{hamricci}
\end{eqnarray}
\vskip 0.5 cm
\noindent where ${\tilde R}_{ik}=R_{ik}-\frac{1}{3}g_{ik}R$ is the trace-free
part of the Ricci tensor. From this expression we see that, not only
shear anisotropies but, also metric anisotropies favour an increasing
in $G(\eta)\rho(\eta)$ with the scale. To give an explicit example,
let us consider as an initial metric to be smoothed-out a locally
homogeneous and anisotropic $SU(2)$-metric g. Following the notation
and the analysis in the paper of Isenberg and Jackson \cite{Jackson},
we can write such a metric and its Ricci-Hamilton evolution, in terms
of a left-invariant one-form basis $\{\theta^a \}$, $a=1,2,3$, on
$SU(2)$ as
\be
g=A(\eta)(\theta^1)^2+B(\eta)(\theta^2)^2+C(\eta)(\theta^3)^2,
\ee
where $A$, $B$, $C$ are scale ($\eta$)-dependent variables. With
respect to this parameterization, the scalar curvature is given by
\be
R(\eta)=\frac{1}{2}\left[ [A^2-(B-C)^2]+[B^2-(A-C)^2]+ [C^2-(A-B)^2]
\right].
\ee
While the squared trace-free part of the Ricci tensor is given by
\begin{eqnarray}
||{\tilde Ric}||^2 &= & \frac{1}{6}
\left[ [A^2-(B-C)^2]^2+[B^2-(A-C)^2]^2+
[C^2-(A-B)^2]^2 \right] \\ &- & \frac{1}{6}\left[
[A^2-(B-C)^2][B^2-(A-C)^2]+ [A^2-(B-C)^2][C^2-(A-B)^2] \right]
\nonumber\\ & +&\frac{1}{6} [B^2-(A-C)^2][C^2-(A-B)^2]. \nonumber
\label{ricciani}
\end{eqnarray}

The Ricci-Hamilton flow for this metric g exponentially converges to
the fixed point $A=B=C=1$, with the normalization $ABC=1$, and with
$A(\eta)\geq B(\eta)\geq C(\eta)$ for all $\eta$
\cite{Jackson}. From the above expression for $R(\eta)$ it follows that
$R(\eta)$ monotonically increases from its initial value $R(\eta=0)$
and exponentially approaches
\be
\lim_{\eta\to\infty}R(\eta)=\frac{3}{2}.
\ee
This increase is generated by the exponential damping of the
anisotropic part of the Ricci tensor (\ref{ricciani}) which is
smoothed by the Ricci-Hamilton flow and, roughly speaking, is
redistributed uniformly in the form of scalar curvature.

Notice that if the anisotropy is large, the actual increase >from
$R(\eta=0)$ to $\lim_{\eta\to\infty}R(\eta)=\bar{R}$ may be quite
significant.  For instance, if for $\eta=0$, we have $A(\eta=0)=1$,
$B(\eta=0)=2$, $C(\eta=0)=1/2$, we get $R(\eta=0)=7/8$, while $\bar{R}
= 3/2$. The renormalized scalar curvature will have increased by
nearly as much as $70\%$ of its original value.
\vskip 0.5 cm

Thus, if we smooth-out the initial data set $(\Sigma\simeq S^3, g,
K,\rho,{\bf J})$, with $g$ the above $SU(2)$-metric,
$K_{ab}=\frac{1}{3}g_{ab}k$, with $k=const.$, $\vec{J}=0$, and with
the matter density $\rho$ such that the Hamiltonian constraint holds,
we get that $G(\eta)\rho(\eta)$ monotonically increases as
$\eta\to\infty$, exponentially approaching a fixed value.
\vskip 0.5 cm
Under the above simplifying assumptions concerning the shear and the
rate of volume expansion, we get
\be
8{\pi}\bar{G}\bar{\rho}=8{\pi}G<\rho>_o+\bar{R} -<R>_o
\ee
\vskip 0.5 cm
\noindent and, as remarked above, the correction term
$\bar{R} -<R>_o$ may be quite large, of {\it the same order of
magnitude as the naive average} $8{\pi}G<\rho>_o$.
\vskip 0.5 cm
More generally, we can easily get an expression for the variation of
$G(\eta)<\rho(\eta)>$ as $\eta$ is increased from a given scale, say
from $\eta_o$ to a larger scale $\eta=\eta_o+\delta\eta$.\par  From
equation (\ref{hamricci}) we immediately get
\begin{eqnarray}
8{\pi}G(\eta)<\rho(\eta)>&=&8{\pi}G(\eta_o)<\rho(\eta_o)>+ 2<{\tilde
R}^{ik}{\tilde R}_{ik}>_{(\eta_o)}\delta\eta \\ &+&
\frac{1}{3}\left(<R>^2-<R^2> \right)_{(\eta_o)}\delta\eta + {\cal
O}(\delta\eta^2).\nonumber
\label{scalevar}
\end{eqnarray}

\noindent With our simplifying assumptions, that there is no shear and
that the rate of volume expansion is spatially constant, we can
exploit the Hamiltonian constraint and rewrite (\ref{scalevar}) as
\vskip 0.5 cm
\begin{eqnarray}
G(\eta)<\rho(\eta)>&= & G(\eta_o)<\rho(\eta_o)>+
\frac{1}{4\pi}
<{\tilde R}^{ik}{\tilde R}_{ik}>_{(\eta_o)}\delta\eta \\ &- &
\frac{8}{3}{\pi}G^2(\eta_o)<\rho(\eta_o)>^2
\left(
\frac{<\rho^2(\eta_o)>-<\rho(\eta_o)>^2}{<\rho(\eta_o)>^2}
\right)\delta\eta+{\cal O}(\delta\eta^2).\nonumber
\label{seesaw}
\end{eqnarray}
\vskip 0.5 cm
 \noindent Thus, in leading order, a large density contrast
$\frac{<\rho^2(\eta_o)>-<\rho(\eta_o)>^2}{<\rho(\eta_o)>^2}$, at a
given observational scale $\eta_o$, tends to reduce the value of
$G(\eta)<\rho(\eta)>$ as the scale of observation is increased. This
reduction effect should dominate on local scales where large
inhomogeneities are present. On sufficiently large scales $\eta_o$,
cosmological data suggest that the density contrast tends to decrease,
and in such a case, as the scale of observation is further increased,
equation (\ref{seesaw}) shows that $G(\eta)<\rho(\eta)>$ now tends to
become larger, with deviations from the original value driven by
anisotropies in the curvature.

Thus, the behavior of the product $G(\eta)<\rho(\eta)>$ under a
variation of the observational scale is related to a competition
between anisotropy and density contrast, and it may be of significance
in the correct interpretation of recent cosmological data.  In
particular, a sufficiently large spatial anisotropy (even maintaining
local homogeneity) may increase $G(\eta)<\rho(\eta)>$ in a significant
way as $\eta$ is increased.
\vskip 0.5 cm

\subsection{The choice of the initial data set}
\label{dataset}

The general picture arising from the above analysis is that we pick up
an appropriate initial data set which, when evolved, gives rise to the
{\it real} space-time. The description of this data set and of the
resulting space-time is too detailed for being of relevance to
cosmology.  Intuitively, one would like to eliminate somehow all the
unwanted (coupled) fluctuations of matter and space-time geometry on
small scales, and thus extract the effective dynamics capturing the
global dynamics of the original space-time. The possibility of
actually implementing such an approach is strongly limited by the fact
that we do not know {\it a priori} the structure of the space-time we
are dealing with.  But we may alternatively decide to handle the
unwanted fluctuations at the level of data sets, since the time
evolution of the initial data set for the Einstein equations is
actually determined by the very constraints which that data set has to
satisfy.  As we have seen above, this can be done quite effectively in
the framework of the Renormalization Group which is naturally well
suited to this purpose.  However, and here we come to the point we
wish to make clear, {\it different initial data sets giving rise to
the same inhomogeneous and anisotropic space-time, may yield smoothed
data set giving rise to different FLRW space-times}.  In other words,
{\bf the renormalization procedure and the dynamics do not commute}.
The dynamics gives rise to the crossover between different FLRW
space-times, or more generally, between different renormalized models
of the same original irregular space-time.
\vskip 0.5 cm
This situation is in fact not so paradoxical as it may seem. From a
thermodynamical point of view, we have seen that one of the members of
the initial data set, namely the three-metric $g$, plays the r\^{o}le
of a temperature. Thus, by varying the metric, one can move through
the possible {\it pure phases} of the thermodynamical system
considered.  In this sense, a real, locally inhomogeneous universe, is
to be considered as akin to a generic complex thermodynamical system.
The possible locally homogeneous cosmological models, arising from it
by suitable choices of initial data set to smooth-out, correspond to
its {\it distinct pure phases}. The resulting dynamics yields, in an
analogy with common statistical system, a {\it dynamical crossover
between different pure phases}.  This makes accessible in cosmology
too, the whole subject of {\it critical phenomena} with a plethora of
interesting consequences.  Critical phenomena are always manifested
macroscopically, as phase transitions are {\it collective} phenomena
in their nature. This aspect may turn out to be of importance for the
study of structure formation and clustering in the universe.
\vskip 0.5 cm

The issue that now needs consideration concerns the existence of
natural initial data sets, in the real lumpy universe, with respect to
which the above Renormalization Group smoothing can be implemented.
As remarked previously, the formalism is particularly well suited to a
data set supported on slices of constant extrinsic time, ({\it viz.},
spatially constant rate of volume expansion, $k$). This is so simply
because the Ricci-Hamilton flow, characterizing the scale-dependence
in the fluctuations of geometry, is most conveniently normalized to
preserve volume. However, depending on the particular geometry we wish
to smooth, different normalizations can be envisaged too
\cite{Jackson}, and the relevance of constant extrinsic time
initial data set can be simply traced back to their importance in the
standard analysis of the initial value problem in relativistic
cosmology.
\vskip 0.5 cm

A suitable slice of a frame comoving with matter is another {\it
natural} choice that almost immediately comes about. However, in
general, we cannot choose surfaces orthogonal to the cosmic fluid flow
lines, since if rotation is present, they do not exist. A more proper
choice would be to select {\it the surface of constant matter density}
as the surfaces of constant time, (see \cite{Giorgetto} and references
quoted therein). It has been argued recently that this choice is
rather optimal. In such a slicing \cite{Giorgetto}, and in the {\it
observable domain}, the spacetime metric takes the form
\be
g^{(4)}=-A^2(x^{\alpha})dt\otimes dt+S^2(t,x^h)
f_{ij}(t,x^h)dx^i\otimes dx^j,
\ee
where $\alpha=1,2,3,4$, $i$,$j=1,2,3$. From an observative point of
view, the function $A^2(x^{\alpha})$ is nearly constant since the
acceleration of the corresponding timelike congruence is small;
$S^2(t,x^h)$ is nearly independent of time since the expansion of the
congruence is almost spatially constant, and finally also $f_{ij}(t,
x^h)$ is nearly independent of time, because the shear is nearly zero.
These remarks suggest that the slices of constant time characterized
in this way are, at least in our observative domain, nearly spaces of
constant curvature, so that they are the most suitable for
implementing the above smoothing procedure.
\vskip 0.5 cm
In this connection notice that, for an initial data set associated
with such a surface of constant matter density, some of the formulae
of the previous sections are simplified. We refer in particular to the
expression (\ref{seesaw}) providing the scale dependence of
$G(\eta)<\rho(\eta)>$ which now reduces to
\vskip 0.5 cm
\be
G(\eta)<\rho(\eta)>= G(\eta_o)<\rho(\eta_o)>+
\frac{1}{4\pi}
<{\tilde R}^{ik}{\tilde R}_{ik}>_{(\eta_o)}\delta\eta +{\cal
O}(\delta\eta^2).
\ee
\vskip 0.5 cm
Thus, for an initial data set associated with a surface of constant
matter density, the product $G(\eta)\rho(\eta)$ increases with the
averaging scale as we average-out the local anisotropies in the
geometry. And, as remarked in the previous section,
$\overline{G\rho}-G\rho$ can be quite large, of the same order of
magnitude of $G\rho$, if $f(x^h,t)$ is sufficiently spatially
anisotropic.
\vskip 0.5 cm

Finally, one may wish to choose initial data set corresponding to a
frame minimizing the anisotropies in the CMB radiation.  This is
characterized by that unique four-velocity which eliminates the CMB
dipole. This corresponds to a family of observers moving with the
background radiation. This choice has an advantage that it can be
accurately determined by local observations, but it is manifestly not
very suitable to use in the Ricci-Hamilton formalism, as developed
here.

\vskip 1 cm

\section{Concluding remarks}

According to the contents of this paper the key idea on which our
whole analysis rests is that of RG, namely, the involved physics is
that of the running (scale dependence, be it energy or momentum
scales) of the couplings and the relevant quantities, accordingly.
This philosophy is recognized and established in particle physics,
e.g. it is well known that the fine structure constant $\alpha$
measured at low energies is different from the one measured at the LEP
energy scale. In each case it is the presence of ``fluctuations" (of
any kind) that requires a scale dependent redefinition (``dressing")
of the physical parameters which can, in turn, modify them, as well as
the very structure of the theory, in a non-trivial way.  Applications
of the RG to particle physics have usually been in the ultra-violet
limit (e.g. in QED, QCD, GUT) whereas in condensed matter physics they
have been in the infra-red limit, in the study of critical phenomena
and phase transitions.

We have taken this infra-red direction in cosmology.  The application
of the concept of running of the physical quantities, motivated by RG,
appears to be a new important feature in a cosmological setting,
providing (at least) partial explanation of some controversies of
standard cosmology, which we are going to discuss below. Let us first
point out that generally running, also, of cosmological quantities is
as such motivated by the asymptotically free higher derivative quantum
gravity, according to which the gravitational constant is
asymptotically free \cite{Avramidy}.  Taking into account this fact,
of running $G$, one can explore its consequences in the standard FLRW
cosmology (cf. \cite{Juan, Bertolami}).

A word of caution is in order - since we do not have an ultimate
theory of quantum gravity (QG), such approaches to cosmology are not
on a rigorous basis from a theoretical point of view and should rather
be taken as phenomenological. In principle, once we have a valid QG
theory will one be able to directly derive a RG equations for various
cosmological quantities. Although we have taken here a more standard
view in which a split exists between the background (associated with
infra-red effects) and renormalization of fluctuations,
%(usually taken to be associated with ultra-violet effects),
there may well be scales where such a split is not sensible at all.
However, a RG capable of interpolating between the qualitatively
different degrees of freedom in a parameter space of gravity, remains
to be developed which may, after all, as well be possible after QG
theory is within our reach.
\bigskip

One of the major issues in modern cosmology is concerned with the
value of the Hubble constant and the apparent conflict between the
observed age of the universe and the predicted one, in the standard
FLRW model, based on the recent measurements of the Hubble constant.
Namely, the recent measurements of the Hubble constant using the Virgo
cluster (distance $\simeq$ 15 Mpc) strongly support that the Hubble
constant $H_o$ has somewhat larger value $h=0.87\pm0.07$\footnote{$h$
is $H_o$ measured in units of $100\,km/sec/Mpc$}
\cite{Pierce}. At the same time, other distance indicators yield a
systematically smaller value of $H_o$, e.g. around $0.55\pm .08$ using
NGC 5253 at a distance of 4 Mpc \cite{Sandage}, while an analysis of
the gravitational lensing of QSO 0957$+$561 indicates $h=0.50\pm 0.17$
\cite{Rhee}. When one calculates now the age of the universe, using
the larger value of $H_o$, in the FLRW model one runs into a serious
problem, as the predicted age turns out to be too small to accomodate
the measured ages ($\sim$14-18 Gyr) of the globular clusters in our
galaxy \cite{agecl, agecl1}.

Moreover, typical inferred values of the density parameter $\Omega_o=
\rho_o/\rho_o^{crit}$ ($\rho_o$ is the present value of the total
energy density of the universe and $\rho_o^{crit}$ its present
critical energy density, defined as $\rho_o^{crit}=3H_o^2/(8\pi G)$
where $G$ is Newton's gravitational constant) increase correspondingly
with the increasing size of various structures (e.g.
\cite{KolbTurner}).  These measurements can at most account for a
fraction of $\Omega_o$ which, according to the inflationary paradigm,
should be equal to 1.  This in turn is one of the reasons for
postulating the existence of non-baryonic Dark Matter (DM) which is
also required to explain the structure formation.
\bigskip

Various people have since then looked at possible theoretical
alternatives to these DM scenarios, such as e.g. introducing a
cosmological constant in the Einstein equations or {\it ad hoc}
modifications of the usual theory of gravity.  The important point in
this respect may as well be the one addressed in this paper. It is
usually taken for granted that, on large scales, the universe is
described by the FLRW solution. There is no alternative really since
we do not know any solution of Einstein's equations capable of
describing a clumpy universe. Nevertheless, even in the absence of
explicit fine-grained models, we would like to know how in principle,
and when, one could extract a background model from an inhomogeneous
one, such that: $(i)$ they both obey, ``approximately", the Einstein
equations despite the averaging or smoothing involved, and $(ii)$
observational determinations of cosmological parameters ($H_o,\,
\Omega_o$,...) correspond in a sensible way to that mathematical
averaging procedure.  Thus an issue of importance for cosmology
\cite{Matravers}, is the question on what scale is the FLRW model
supposed to describe the universe? Likewise, what averaging scale are
we referring to when we give the value of $\Omega_o$, whose definition
necessarily refers to an idealised, i.e.  smoothed background model.

There has been recently an increased effort in this direction with
some interesting results, as e.g. that the coarse-graining effects
could be non-negligible in the context of affecting the age of the
universe.  For example, \cite{Suto} considered a model, of locally
open (underdense) universe embedded in the spatially flat universe, in
which the expansion rate in our local universe is larger than the
global average. Similar model was considered in \cite{Moffat} where a
local void in the global FLRW model was studied and the inhomogeneity
described by the Lema\^\i tre-Tolman-Bondi solution.  The results
indicate that if we happened to live in such a void, but insisted on
interpreting the observations by the FLRW model, the Hubble constant
measurements could give results depending on the separation of the
source and the observer, providing a possible explanation for the wide
range of their reported values and capable of resolving the
age-of-universe problem.
\bigskip

On the other hand \cite{Juan} studied the QG effects at cosmological
scales (the phenomenon in question is exactly that of quantum
coherence, known from a laboratory to happen on macroscopic scales of
the order of cm), assuming asymptotic freedom of the gravitational
constant and incorporating running $G$, according to the appropriate
RG equations, into FLRW model. Such $G$ takes the value of Newton's
constant $G_N$ at short distances but then slowly rises as distance
increases. However, as mentioned in \cite{Juan}, the RG equations used
there might not be applicable in the infra-red regime studied, but
these concerns were put aside having in mind the absence of any other
available beta functions for QG in the infra-red regime.

One can also approach the averaging problem, modifying the FLRW metric
(or equivalently Cosmological Principle) and the Einstein equations,
by an introduction of a generalized scale factor which depends both on
$t$ and the scale $r$ \cite{Kim}. This introduces the scale dependence
of $(G\rho)$ and running of other cosmological quantities such as
$H_o$, $\Omega_o$, and the age of the universe $t_o$, as functions of
distance scales.
\bigskip

Let us notice that this picture of running of cosmological quantities
comes about naturally in our approach which is physically motivated by
RG, namely, due to increasing of the gravitational constant with
scale, (and possibly increasing amount of DM), as discussed in the
previous section, $\Omega_o$ has effectively increased too. Moreover,
since the scale factor is governed by the scale dependent $(G\rho)$,
it now seems to depend on the scale $r$ as well, i.e. it increases at
any fixed time with the increasing distance.  Consequently, the value
of $H_o$ is not the same everywhere in the observable universe and
depends on a scale where it is measured.  Moreover, since $H_o\sim
1/t_o$ the universe becomes older when its age is estimated on a
smaller scale. This is not to be taken as implying that the age
depends on where one calculates (every observer using the same scale
$r$ at some time $t_1$ will obtain the same age). The key point to
emphasize is that, having in mind the RG arguments and interpretation,
a direct comparison of cosmological quantities makes sense only when
they are measured (or calculated) with respect to the same scale,
since the same quantity can take different values at different scales.

Notice that independently, also Quantum Cosmology advocates, though in
a different context of bubble universes, a possibility that we may
live in a universe in which the value of Hubble constant and the
measured density are different in different places and in our local
neighborhood $\Omega_o$ may well be less than 1 \cite{Linde}.

 This makes our proposal even more promising.

\bigskip

{\bf Acknowledgements}

We wish to thank Juan P\'{e}rez-Mercader and Giancarlo Jug for
interesting discussions on the Renormalization Group.

This work has been supported by MURST, the National Institute for
Nuclear Physics (INFN), and by the EEC-Contract {\it Constrained
Dynamical System}, (Human Capital and Mobility Program)
n.CHRX-CT93-0362.

%%%%%%%%%%%%%%%%%%%%%%%%%%%%%%%%%%%%%%%%%%%%%%

%\newpage

\newpage
{\bf Figure Captions}\par
\vskip  1cm
{\bf Fig. 1} A portion of a minimal geodesic balls covering. The
dotted disks are the $\frac{\epsilon}{2}$-balls which pack a given
region. The larger, undotted disks, represent the $\epsilon$-balls
which provide the covering.
\vskip 1cm
{\bf Fig. 2} We can approximate the average over $\Sigma$ by averaging
over euclidean balls whose Lebesgue measure has been locally weighted
through Puiseux' formula. In this drawing, the
$\frac{\epsilon}{2}$-geodesic balls are represented by curved disks on
$\Sigma$, while the corresponding euclidean balls are correctly
depicted as three-dimensional.
\vskip 1 cm
{\bf Fig. 3} The local average of a function $f$ over a geodesic ball
$B(x,\epsilon_o)$ feels the underlying curvature of the manifold
through Puiseux' formula. In particular, in passing from
$B(x,\epsilon_o)$ to the larger ball $B(x,\epsilon_o+\eta)$ we get
correction terms which depend on the fluctuations in curvature.
\vskip 1 cm
{\bf Fig. 4} The intricate but symmetrical intersection pattern of an
array of $\frac{\epsilon}{2}$-balls in the $2D$-plane. Here, the
$\frac{\epsilon}{2}$-balls are dotted while the overlapping
$\epsilon$-balls, providing the covering, are the dashed circles.  The
solid circles and the arcs of circles describe the intersection
pattern of a $\frac{3}{2}\epsilon$-ball. On a curved manifold of
bounded geometry, the pattern is more complicated and not symmetrical
at all. Nonetheless, one may easily obtain a recursive definition of
blocking by exploiting a partition of unity argument, explained in the
text.
\vskip 1 cm
{\bf Fig. 5} The intersection pattern of a $2{\epsilon}$-ball
$B(x_j,2\epsilon)$ with five $\epsilon$-balls $B(x_h,\epsilon)$. The
$2\epsilon$-ball is represented by a dotted disk, while the
$\epsilon$-balls are represented by solid circles. The white disks
represent the packing of $B(x_j,2\epsilon)$ with
$\frac{\epsilon}{2}$-balls. The standard representation of the
integral over $B(x_j,2\epsilon)$, through the partition of unity
subordinated to the $B(x_h,\epsilon)$-covering, yields the recursive
relation $\psi_{m+1}(j)=\sum_h\psi_m(h)$.
\vskip  1cm
{\bf Fig. 6} Even if the function f is constant over $\Sigma$, its
average values over the geodesic balls, $B(x_1,\epsilon)$ and
$B(x_2,\epsilon)$, are generally uncorrelated owing to the random
fluctuations in the geometry of the balls.
\vskip 1cm
{\bf Fig. 7} The matter distribution $\rho$, $\vec{J}$, the lapse
function $\alpha$, and the shift vector field $\vec{\alpha}$
associated with the istantaneous observers on $\Sigma$. Here,
$\vec{n}$ denotes the four-dimensional unit, future-pointing, normal
to $\Sigma$. The vector field $\alpha\vec{n}+\vec{\alpha}$ connects
the corresponding points in the two infinitesimally close slices
$\Sigma$ and $\Sigma_{\delta t}$.
\vskip 1cm
{\bf Fig. 8} Given a probability law, according to which matter is
distributed at a given length scale (say, the planetary scale), we can
get the corresponding probability distribution obtained by averaging
the matter variables over regions of ever increasing scales. The
resulting {\it effective} Hamiltonians are defined up to additive
constants which affect the renormalized mass density and the
renormalized momentum density. Such indeterminacy can be naturally
removed by enforcing the Hamiltonian and divergence constraints at
each step of the renormalization procedure.
\vskip 1cm
{\bf Fig. 9} Upon enlarging the ball, from $\epsilon$ to
$e^{\gamma}\epsilon$, we alter the average value of the matter
variables $\rho$ and $\vec{J}$, (even if $\rho$ and $\vec{J}$ do not
vary appreciably), since there can be strong curvature fluctuations in
the annulus $B(x,e^{\gamma}\epsilon)\backslash B(x,\epsilon)$.
\vskip 1 cm
{\bf Fig. 10} If we enlarge the ball from $\epsilon$ to
$e^{\eta}\epsilon$, while deforming the metric according to the
Ricci-Hamilton flow (\ref{ham}), (for a {\it parameter time}
$\eta=\gamma$), then the average value remains {\it as constant as it
can be} since (\ref{ham}) smoothes the curvature fluctuations in the
annulus $B(x,e^{\gamma}\epsilon)\backslash B(x,\epsilon)$.
\vskip 1 cm
{\bf Fig. 11} If we blow up the metric of two three-spheres
$S^3_{(1)}$ and $S^3_{(2)}$, (here represented as two-spheres), around
the points $y_1$ and $y_2$ and then join the resulting manifolds, we
obtain the connected sum $S^3_{(1)}\sharp S^3_{(2)}$. According to
this procedure, the neck $S^2\times{\rm I\!R}^1$ inherits a cylindrical
metric up to exponentially small correction terms.
\vskip 1 cm
{\bf Fig. 12} The Ricci-Hamilton flow of $S^3_{(1)}\sharp S^3_{(2)}$
can be expressed in terms of the $\eta$-evolution of the inclusion
maps $\chi_{(\alpha)}$, $\alpha=1,2,3$, and of the known evolution of
the locally homogeneous metrics on $S^3_{(i)}\backslash B(y_i)$ and
$S^2\times{\rm I\!R}^1$.
\vskip 1 cm
{\bf Fig. 13} Different deformations of the initial three-manifold
$\Sigma$ may have quite different fates. If the waist of $\Sigma$ is
increased enough by the deformation, (i.e., if it is rounded), the
Ricci-Hamilton flow will evolve $\Sigma$ toward the round
three-sphere. If the waist is shrinked enough by the initial
deformation, the Ricci-Hamilton flow will evolve $\Sigma$ toward
$S^3_{(1)}\coprod S^3_{(2)}$.  The three-manifolds attracted towards
$S^3_{(1)}\coprod S^3_{(2)}$ define a {\it critical surface}.
\vskip 1 cm
{\bf Fig. 14} Two neighbooring three-manifolds, $(\Sigma, g)$ and
$(\Sigma, g+\delta K)$, with the same volume, evolving under the
Ricci-Hamilton flow toward isometric round three-spheres $(S^3,
g_{can})$ and $(S^3, g^*_{can})$.  In general, $g_{can}$ and
$g^*_{can}$ differ by an infinitesimal diffeomorphism $\phi$ generated
by a vector field $v$, i.e.  $g^*_{can}=g_{can}+\delta\left(
L_vg_{can}-\frac{2}{3}g_{can}
\nabla_iv^i   \right).$
By exploiting the freedom in choosing the shift vector field
$\vec{\alpha}$, we can get rid of $\phi$.
\vskip 1 cm
{\bf Fig. 15} The time evolution of an inhomogeneous and anisotropic
initial data set and its Ricci-Hamilton renormalized counterpart.
Notice, that $({\bar M}^4, {\bar g}^{(4)})$ depends in a {\it sensible
way} on the particular initial data set $(\Sigma, g, K,\rho,\vec{J})$
which is renormalized according to the Ricci-Hamilton flow. Different
initial data sets, corresponding to the same inhomogeneous and
anisotropic space-time $(M^4,g^{(4)})$, may generate {\it distinct}
renormalized model space-times $({\bar M}^4, {\bar g}^{(4)})$. This
seemingly paradoxical situation has a natural explanation in terms of
the RG approach.

\end{document}